\pdfoutput=1
\documentclass[aps,prd,floatfix,amsmath,amssymb,onecolumn,groupedaddress,11pt]{revtex4}
\usepackage{graphicx}
\usepackage{amssymb}
\usepackage{epstopdf}
\usepackage{subfigure}
\usepackage{array}
\usepackage{color}
\usepackage{pdfsync}
\usepackage{amsmath}
\usepackage{hyperref}
\newcommand{\st}[0]{\mathcal{S}_{3}}

\begin{document} 

\title{Generating the Cabibbo Angle in Flavored Gauge Mediation Models with Discrete Non-Abelian Symmetries}
\author{Lisa L.~Everett}
\email{leverett@wisc.edu} 
\author{Todd S.~Garon}
\email{tgaron@wisc.edu}  
\author{Ariel B.~Rock}
\email{arock3@wisc.edu}
\affiliation{Department of Physics, University of Wisconsin-Madison, Madison, WI 53706}
\date{\today}

\begin{abstract}
We explore the generation of fermion masses and quark mixing angles within flavored gauge mediation models of supersymmetry breaking in which the Higgs and messenger doublets are connected by a discrete non-Abelian symmetry.  In this framework, the Higgs-messenger symmetry, which we take for concreteness to be the discrete group $\mathcal{S}_3$, also plays the role of a (partial) family symmetry.  We investigate mechanisms for generating mass hierarchies for the lighter quark generations and generating the Cabibbo angle within this class of models.  While perturbations of the couplings at the renormalizable level do not lead to phenomenologically viable quark mixing parameters, we show that the Cabibbo angle can be generated via couplings at the nonrenormalizable level and explore the phenomenological implications of this scenario.

%We explore the model-building and phenomenology of flavored gauge mediation models of supersymmetry breaking in which the electroweak Higgs doublets and the $SU(2)$ messenger doublets are connected by a discrete non-Abelian symmetry.  The embedding of the Higgs and messenger fields into representations of this non-Abelian Higgs-messenger symmetry results in specific relations between the Standard Model Yukawa couplings and the messenger-matter Yukawa interactions.  Taking the concrete example of an $\mathcal{S}_3$ Higgs-messenger symmetry, we demonstrate that while the minimal implementation of this scenario suffers from a severe $\mu/B_\mu$ problem that is well-known from ordinary gauge mediation,  expanding the Higgs-messenger field content allows for the possibility that $\mu$ and $B_\mu$ can be separately tuned, allowing for the possibility of phenomenologically viable models of the soft supersymmetry breaking terms.  We construct toy examples of this type that are consistent with the observed 125 GeV Higgs boson mass.  

\end{abstract}
\maketitle 
%\tableofcontents

\section{Introduction}

Gauge-mediated supersymmetry breaking \cite{gauge1,gauge2,gauge3,Giudice:1998bp} provides an elegant framework for the generation of the soft supersymmetry breaking sector of the minimal supersymmetric standard model (MSSM) and its extensions.  In its minimal incarnation within the MSSM, this framework is well known to be severely constrained by the LHC Higgs data \cite{Draper:2011aa,Arbey:2011ab,Ajaib:2012vc}, as it predicts negligibly small scalar trilinear couplings ($A$ terms) at the messenger scale. One way to circumvent this issue is to consider non-minimal versions of gauge mediation, in which the messenger fields have direct (renormalizable) couplings to the MSSM fields \cite{gauge3,Giudice:1998bp,Chacko:2001km,Shadmi:2011hs,Evans:2011bea, Evans:2011uq,Evans:2012hg,Kang:2012ra,Craig:2012xp,Albaid:2012qk,Abdullah:2012tq,Perez:2012mj,Byakti:2013ti,Evans:2013kxa,Calibbi:2013mka,Evans:2015swa,Galon:2013jba,Fischler:2013tva,Calibbi:2014yha,Ierushalmi:2016axs}.  
A compelling set of examples within this broad category are ``flavored gauge mediation'' models~\cite{Shadmi:2011hs,Abdullah:2012tq,Perez:2012mj,Byakti:2013ti,Evans:2013kxa,Calibbi:2013mka,Evans:2015swa,Galon:2013jba,Fischler:2013tva,Calibbi:2014yha,Ierushalmi:2016axs},  
for which there is nontrivial mixing of the $SU(2)_L$ messenger doublets and the electroweak Higgs fields of the MSSM.  Flavored gauge mediation models allow for the generation of nontrivial $A$ terms at the messenger scale, thus alleviating the Higgs mass problem of minimal gauge mediation in the MSSM.  

Flavored gauge mediation also provides an intriguing setting for exploring the flavor puzzle of the Standard Model (SM). Since the electroweak Higgs doublets mix with the messenger doublets, the generation of the needed messenger Yukawa couplings is tied together with the generation of the Yukawa couplings of the quarks and leptons.  In such flavored gauge mediation models, the underlying Higgs-messenger symmetry that controls the mixing of the messenger doublets and the Higgs doublets can also play a role as (part of) the family symmetry that controls the generation of the fermion masses of the SM, if the MSSM matter is also nontrivially charged under the Higgs-messenger symmetry.  While non-minimal flavor violation can then result, opening the door to the supersymmetric flavor problem ~\cite{Gabbiani:1988rb,Hagelin:1992tc,Gabbiani:1996hi,Raz:2002zx}, it has been shown that flavor-violating effects in these models can often be more strongly suppressed than naive estimates might indicate~\cite{Calibbi:2014yha,Ierushalmi:2016axs}.

An intriguing possibility is that the Higgs-messenger symmetry is a discrete non-Abelian symmetry (for examples based on $U(1)$ symmetries, see e.g.~\cite{Ierushalmi:2016axs}). This idea was first studied for the case of a $\mathcal{S}_3$ Higgs-messenger symmetry for two families \cite{Perez:2012mj}, and later extended to three families \cite{Everett:2016meb,Everett:2018wrn}. Discrete non-Abelian symmetries provide rigid constraints on these models, with implications not only for the flavor puzzle but also for the well-known $\mu/B_\mu$ problem of gauge mediation \cite{Dvali:1996cu,Giudice:2007ca} (see \cite{Giudice:1998bp,Polonsky:1999qd} for reviews).  In  \cite{Everett:2016meb,Everett:2018wrn}, we showed that three-family models based on $\mathcal{S}_3$ require an expanded Higgs-messenger sector that results in two pairs of heavy messenger doublets, as well as the pair of light doublets that are to be identified as the Higgs fields $H_{u,d}$ of the MSSM.  For the case of interest in which the MSSM matter superfields also are embedded within $\mathcal{S}_3$ representations, various scenarios can be constructed with one heavy and two massless generations. Such scenarios can also be consistent with the Higgs mass constraints for squark masses in the $5-6$ TeV range.  

In this paper, we build upon \cite{Everett:2016meb,Everett:2018wrn} to explore the implications for the flavor puzzle within these scenarios in greater detail.  Our focus in particular is on achieving a mass hierarchy for the lighter generations of quarks and charged leptons, and to obtain a reasonable value for the Cabibbo mixing angle of the quark sector (here we will ignore the issue of neutrino mass generation, and return to this question in future work).  We find that while specific perturbations of the Yukawa couplings associated with the renormalizable superpotential interactions do not in general lead to the appropriate mixing of the first and second generations, the Cabibbo angle can be generated at the nonrenormalizable level, with corresponding implications for the mass spectrum of the theory.

The structure of this work is as follows.  We begin with an overview of the discrete non-Abelian Higgs-messenger symmetry and the resulting model structure, focusing on the case of the discrete group $\mathcal{S}_3$, as in  \cite{Perez:2012mj,Everett:2016meb,Everett:2018wrn}.  We present the model and discuss the generation of masses and mixing angles among the first and second quark families. The phenomenological implications are then discussed.  Finally, we present our summary and conclusions.

\section{Theoretical Overview}
\label{backgroundsection}

\noindent In the class of models we consider, the Higgs-messenger symmetry is taken to be $\mathcal{S}_3$, the permutation group on three objects. Its associated group theory can be found in many references (see e.g.~\cite{Perez:2012mj}). Here we summarize its most salient features for our study.   

$\mathcal{S}_3$ contains three irreducible representations: the singlet $\mathbf{1}$, a one-dimensional representation $\mathbf{1}^\prime$, and a doublet, $\mathbf{2}$. The tensor products involving the doublets are 
\begin{eqnarray}
\mathbf{1}\otimes \mathbf{2}=\mathbf{2}, \qquad \mathbf{1}^\prime \otimes  \mathbf{2}=\mathbf{2}, \qquad  \mathbf{2}\otimes \mathbf{2}=\mathbf{1}\oplus \mathbf{1}^\prime\oplus\mathbf{2}.
\end{eqnarray}
As in \cite{Perez:2012mj}, we use a group presentation such that the singlet representations obtained from the tensor products of either two doublets or three doublets are given by:
\begin{eqnarray}
(\mathbf{2} \otimes \mathbf{2})_\mathbf{1}&=& \left [\left (\begin{array}{c} a_1 \\ a_2 \end{array} \right ) \otimes  \left (\begin{array}{c} b_1 \\ b_2 \end{array} \right ) \right ]_\mathbf{1} = a_1 b_2 +a_2b_1. \nonumber \\
(\mathbf{2} \otimes \mathbf{2}\otimes \mathbf{2})_\mathbf{1}&=& \left [\left (\begin{array}{c} a_1 \\ a_2 \end{array} \right ) \otimes  \left (\begin{array}{c} b_1 \\ b_2 \end{array} \right )\otimes  \left (\begin{array}{c} c_1 \\ c_2 \end{array} \right ) \right ]_\mathbf{1} =a_1b_1c_1+a_2b_2 c_2.
\label{cg}
\end{eqnarray}
Here all fields are taken for simplicity to be either $\mathbf{1}$ or $\mathbf{2}$ representations, such that Eq.~(\ref{cg}) provides us with the relations needed to construct $\mathcal{S}_3$ invariants.  

In this scenario, the Higgs-messenger sector consists of the following collection of chiral superfields that have specific transformation properties with respect to the $\mathcal{S}_3$ symmetry:
\begin{eqnarray}
\mathcal{H}_{u}= \left (\begin{array}{c} \mathcal{H}_{u1}\\ \mathcal{H}_{u2}\\ \mathcal{H}_{u3}\end{array} \right ) = \mathcal{R}_u \left (\begin{array}{c} H_u\\M_{u1} \\  M_{u2} \end{array} \right ), \qquad  \mathcal{H}_{d}= \left (\begin{array}{c} \mathcal{H}_{d1}\\ \mathcal{H}_{d2}\\  \mathcal{H}_{d3} \end{array} \right ) = \mathcal{R}_d \left (\begin{array}{c} H_d\\M_{d1}  \\ M_{d2} \end{array} \right ),
\label{higgs_s3}
\end{eqnarray}
in which the electroweak Higgs fields are denoted by $H_{u,d}$ the $SU(2)$ doublet messengers are given by $M_{ui,di}$ ($i=1,2$), and $\mathcal{R}_{u/d}$ are rotation matrices that are obtained upon diagonalizing the mass matrices of the Higgs/doublet messenger sector of the theory.  The individual components $\mathcal{H}_{ui,di}$ are given by  $\mathcal{S}_3$ doublets (denoted by $\mathcal{H}^{(2)}_{u,d}$) and $\mathcal{S}_3$ singlets (denoted by $\mathcal{H}^{(1)}_{u,d}$). Note that two sets of messenger doublets are included; this is the minimal set needed to accommodate the constraints of the $\mu/B_\mu$ problem.  The theory also includes $SU(3)$ triplet messengers, which are taken to be $\mathcal{S}_3$ singlets, are denoted by $T_{ui,di}$ ($i=1,2$). The SM charges of the $T_{ui,di}$ and the messenger doublets $M_{ui,di}$ are such that together they form two vectorlike pairs of $\mathbf{5}$, $\overline{\mathbf{5}}$ representations of $SU(5)$.  

The model also includes two supersymmetry breaking fields: the $\mathcal{S}_3$ doublet, $X_H$, which couples to the messenger-Higgs fields, and a $\mathcal{S}_3$ singlet chiral superfield $X_T$ that couples only to $T_{ui,di}$ via the superpotential coupling $\lambda_T X_T T_{ui} T_{di}$.  It is further assumed that the triplet messengers and $X_T$ do not have renormalizable couplings to the messenger doublets or the MSSM fields, as needed to avoid rapid proton decay. This typically requires additional symmetries, but this is not difficult to implement in a concrete scenario; what is more difficult is to embed this scenario within a fully grand unified theory. We defer that question to future work.

%This field content and the relevant $\mathcal{S}_3$ charges are summarized in Table~\ref{tab:11}. 
%\begin{table}[htbp]
%   \centering
%    \begin{tabular}{c|cccccc|cc}
%     & $\mathcal{H}_u^{(2)}$&$\mathcal{H}_u^{(1)}$ & $\mathcal{H}_d^{(2)}$& $\mathcal{H}_d^{(1)}$ & $T_{ui}$ & $T_{di}$   &$X_H$ & $X_T$\\
%    \hline
%    $\mathcal{S}_3$ &$\mathbf 2 $& $\mathbf 1$& $\mathbf 2 $& $\mathbf 1 $ & $\mathbf 1 $ & $\mathbf 1 $ &\textbf 2 & $\mathbf 1 $\\
%    %$Z_n$& 
%   \end{tabular}
%   \caption{The $\mathcal{S}_3$ charges for the extended Higgs-messenger model described in this section.}
%   \label{tab:11}
%\end{table} 
As discussed in \cite{Everett:2016meb,Everett:2018wrn}, the superpotential couplings of $X_H$ to the Higgs-messenger sector are given by
\begin{eqnarray}
W_H= \lambda X_H\mathcal{H}^{(2)}_u \mathcal{H}^{(2)}_d+\lambda' X_H \mathcal{H}^{(1)}_u \mathcal{H}^{(2)}_d+\lambda''X_H \mathcal{H}^{(2)}_u \mathcal{H}^{(1)}_d +  \kappa M \mathcal{H}^{(2)}_u \mathcal{H}^{(2)}_d+  \kappa^\prime M \mathcal{H}^{(1)}_u \mathcal{H}^{(1)}_d.
\end{eqnarray}
All couplings are taken for simplicity to be real.  The vacuum expectation value (vev) of the supersymmetry-breaking field $X_H$ is parametrized by
\begin{eqnarray}
\langle \lambda X_H \rangle = M  \left (\begin{array}{c} \sin\phi \\ \cos\phi \end{array} \right ) +\theta^2 F  \left (\begin{array}{c} \sin\xi \\ \cos\xi \end{array} \right ),
\label{deltavev}
\end{eqnarray}
in which $\phi$ and $\xi$ characterize the vev directions of the scalar and $F$ components, respectively, and we take $F\ll M^2$ for simplicity.  After symmetry breaking, the effective superpotential is given by
\begin{eqnarray}
W_H&\equiv& \mathcal H_u^T\mathbb{M} \mathcal H_d+\theta^2 \mathcal{H}_u^T \mathbb{F} \mathcal{H}_d \nonumber \\ &=&  M \mathcal H_u^T \left(\begin{matrix}\sin\phi&\kappa &\epsilon'\cos\phi\\ \kappa &\cos\phi &\epsilon'\sin\phi\\\epsilon''\cos\phi &\epsilon'' \sin\phi &\kappa^\prime \end{matrix}\right)\mathcal H_d  + \theta^2 F \mathcal H_u^T\left(\begin{matrix}\sin\xi&0&\epsilon'\cos\xi\\ 0 &\cos\xi &\epsilon'\sin\xi\\\epsilon''\cos\xi &\epsilon'' \sin\xi & 0 \end{matrix}\right)\mathcal H_d, \nonumber
\label{extendedS3}
\end{eqnarray}
in which $\epsilon'=\lambda'/\lambda$, $\epsilon''=\lambda''/\lambda$, and the quantities $\mathcal{H}_{u,d}$ are now given by
\begin{eqnarray}
\mathcal{H}_u = \left (\begin{array}{c} (\mathcal{H}^{(2)}_u)_1 \\ (\mathcal{H}^{(2)}_u)_2 \\ \mathcal{H}^{(1)}_u \end{array} \right ), \qquad \mathcal{H}_d = \left (\begin{array}{c} (\mathcal{H}^{(2)}_d)_1 \\ (\mathcal{H}^{(2)}_d)_2 \\ \mathcal{H}^{(1)}_d \end{array} \right ).
\label{ourmodeldecomp}
\end{eqnarray}
Here we set $\epsilon'' =\epsilon$, such that $\mathbb{M}$ and $\mathbb{F}$ are symmetric matrices, and set $\epsilon'=1$.   The next step \cite{Perez:2012mj} is to impose $[\mathbb{M},\mathbb{F}]=0$, which yields $\kappa^\prime = \kappa = \sin(\phi-\xi)/(\cos \xi-\sin\xi)$, where $\xi\neq \pi/4$.  

With these constraints, a viable solution with a distinct hierarchy of eigenvalues for both $\mathbb{M}$ and $\mathbb{F}$ can then be obtained.  This distinct hierarchy is needed for separate fine-tunings of the $\mu$ and $b$ parameters, as well as for a clean separation in mass scales between the electroweak Higgs doublets and the doublet messenger fields.  The solution occurs in the limit in which $\xi\rightarrow -\pi/4$ and $\phi\neq \xi$, with a small detuning between $\phi$ and $\xi \simeq -\pi/4$ that controls the size of the resulting $\mu$ term.   In this limit, the matrices $\mathcal{R}_{u,d}$ are given to leading order by
\begin{eqnarray}
\mathcal{R}_{u,d}= \left ( \begin{array}{ccc} \frac{1}{\sqrt{3}} & \mp \frac{1}{2} \left (1+\frac{1}{\sqrt{3}} \right) & \frac{1}{2} \left (1-\frac{1}{\sqrt{3}} \right) \\  \frac{1}{\sqrt{3}} & \pm \frac{1}{2} \left (1-\frac{1}{\sqrt{3}} \right) & -\frac{1}{2} \left (1+\frac{1}{\sqrt{3}} \right) 
\\  \frac{1}{\sqrt{3}} &  \pm \frac{1}{\sqrt{3}} &  \frac{1}{\sqrt{3}} \end{array} \right ).
\label{rotationmatrices}
\end{eqnarray}
Note that the trimaximal vector is associated with the light eigenstate, which is precisely the state that corresponds to the electroweak doublets $H_{u,d}$.  (More precisely, the eigenvalues corresponding to this light eigenstate are $\mu\ll M$ for the case of $\mathbb{M}$, and $b\ll F$ for the case of $\mathbb{F}$.  The heavy states in this limit have equal masses that are proportional to $M$.

\section{Fermion Masses: Renormalizable Couplings}

As studied in \cite{Everett:2018wrn}, a key assumption of this scenario is that the three generations of SM quarks and leptons are embedded into doublet and singlet represenations of $\mathcal{S}_3$. The charge assignments for the fields in the theory is summarized in Table~\ref{tab:12}.
\begin{table}[htbp]
   \centering
    \begin{tabular}{c|cccc|cccccccccc|c}
  & $\mathcal{H}_u^{(2)}$&$\mathcal{H}_u^{(1)}$ & $\mathcal{H}_d^{(2)}$& $\mathcal{H}_d^{(1)}$  & $Q_{\mathbf 2}$ &$Q_{\mathbf 1}$& $\bar u_{\mathbf 2}$ & $\bar u_{\mathbf 1 }$&$\bar d_{\mathbf 2}$& $\bar d_{\mathbf 1}$ & $L_{\mathbf{2}}$ & $L_{\mathbf{1}}$ & $\bar{e}_{\mathbf{2}}$ & $\bar{e}_{\mathbf{1}}$&$X_H$\\
    \hline
    $\mathcal{S}_3$ &$\mathbf 2 $& $\mathbf 1$& $\mathbf 2 $& $\mathbf 1 $  & $\mathbf 2 $&$\mathbf 1$  & $\mathbf 2 $&$\mathbf 1$  & $\mathbf 2 $&$\mathbf 1$ & $\mathbf 2 $&$\mathbf 1$  & $\mathbf 2 $&$\mathbf 1$ &\textbf 2\\
    %$Z_n$& 
 % $\mathcal Z_n$ &$\omega^3 $& $\omega^4 $&  1& 1 &$\omega$& $\omega^{-4}$ & $\omega^{-3}$&$\omega$ &$\omega^{6} $& $\omega $
   \end{tabular}
   \caption{Charges for an $\st$ model of the Higgs-messenger fields and the MSSM matter fields. Here the $SU(3)$ triplet messengers and the associated $X_T$ field are not displayed for simplicity.}
   \label{tab:12}
\end{table} 

The renormalizable superpotential Yukawa couplings of the MSSM matter fields and the Higgs-messenger fields, for example for the up quarks, are given by 
\begin{eqnarray}
W^{(u)}= y_u\big[Q_{\mathbf 2}  \bar u_{\mathbf 2}  \mathcal{H}^{(2)}_u+\beta_1Q_{\mathbf 2}  \bar u_{\mathbf 2} \mathcal{H}^{(1)}_u + \beta_2 Q_{\mathbf 2}  \bar u_{\mathbf 1}  \mathcal{H}^{(2)}_u +\beta_3 Q_{\mathbf 1}  \bar u_{\mathbf 2}  \mathcal{H}^{(2)}_u+ \beta_4Q_{\mathbf 1}  \bar u_{\mathbf 1}  \mathcal{H}^{(1)}_u\big],
%&=y_uQ^T\left( \begin{matrix} \Hu{\mathbf21}&\beta_1\Hu{\mathbf1}&\beta_2\Hu{\mathbf22}\\ \beta_1 \mathcal{H}^{(1)}_u& \Hu{\mathbf22}& %\beta_2\Hu{\mathbf21}\\ \beta_3\Hu{\mathbf22}& \beta_3\Hu{\mathbf21}&\beta_4 \mathcal{H}^{(1)}_u\end{matrix}\right)\bar u, \label{UpYukawas}
\label{wu}
\end{eqnarray}
in which the $\beta_i$ are arbitrary coefficients in the absence of further model structure. We note that here we will take them to be real, for simplicity \footnote{The assumption of real $\beta_i$ clearly has an effect on options for generating the CKM quark mixing phase angle $\delta_{\rm CKM}$. We defer the detailed discussion of generating a viable $\delta_{\rm CKM}$ for a future study.}.   In the basis given by $Q=(Q_{\mathbf 2} ,Q_{\mathbf 1})^T$ and $\overline{u}=(\overline{u}_{\mathbf 2} ,\overline{u}_{\mathbf 1})^T$, these couplings can be expressed in matrix form as  \footnote{Eq.~(\ref{UpYukawas}) and its generalizations to other charged SM fermions correct a typo in the corresponding expressions for the renormalizable superpotential in \cite{Everett:2016meb}, for which there was an incorrect interchange of $\beta_1$ and $\beta_2$; these expressions are correct in \cite{Everett:2018wrn}.}:
\begin{eqnarray}
W^{(u)}=y_uQ^T\left( \begin{matrix} \mathcal{H}^{(2)}_{u1}&\beta_1\mathcal{H}^{(1)}_{u}&\beta_2 \mathcal{H}^{(2)}_{u2}\\ \beta_1 \mathcal{H}^{(1)}_u& \mathcal{H}^{(2)}_{u2}& \beta_2\mathcal{H}^{(2)}_{u1}\\ \beta_3\mathcal{H}^{(2)}_{u2}& \beta_3\mathcal{H}^{(2)}_{u1}&\beta_4 \mathcal{H}^{(1)}_u\end{matrix}\right)\bar u. \label{UpYukawas}
\end{eqnarray}
Analogous coupling matrices would hold in the down quark and charged lepton sectors, with the replacements   $\beta_i\rightarrow \beta_{di},\beta_{ei}$.

Using Eq.~(\ref{ourmodeldecomp}) and Eq.~(\ref{rotationmatrices}) in Eq.~(\ref{UpYukawas}), it is straightforward to see that the SM up quark sector Yukawa couplings are given by
\begin{equation}
Y_u =  \frac{y_u}{\sqrt{3}} \left (\begin{array}{ccc} 1 & \beta_1 & \beta_2 \\ \beta_1 & 1 & \beta_2 \\ \beta_3 & \beta_3 & \beta_4           \end{array} \right ),
\label{eq:yu}
\end{equation}
and the messenger Yukawa couplings $Y^\prime_{u1}$ and $Y^\prime_{u2}$ take the form
\begin{equation}
Y^\prime_{u1}=y_u \left (\begin{array}{ccc} -\frac{1}{2}-\frac{1}{2\sqrt{3}} & \frac{\beta_1}{\sqrt{3}} & \;\; \frac{\beta_2}{2} - \frac{\beta_2}{2\sqrt{3}} \\  \frac{\beta_1}{\sqrt{3}} & \;\; \frac{1}{2}-\frac{1}{2\sqrt{3}} & -\frac{\beta_2}{2} - \frac{\beta_2}{2\sqrt{3}} \\ \;\; \frac{\beta_3}{2} - \frac{\beta_3}{2\sqrt{3}} & -\frac{\beta_3}{2} - \frac{\beta_3}{2\sqrt{3}} & \frac{\beta_4
}{\sqrt{3}}
\end{array} \right )
\end{equation}
\begin{equation}
\;\; Y^\prime_2=y_{u2} \left (\begin{array}{ccc} \;\; \frac{1}{2}-\frac{1}{2\sqrt{3}} & \frac{\beta_1}{\sqrt{3}} &  -\frac{\beta_2}{2} - \frac{\beta_2}{2\sqrt{3}} \\  \frac{\beta_1}{\sqrt{3}} & -\frac{1}{2}-\frac{1}{2\sqrt{3}} & \;\; \frac{\beta_2}{2} - \frac{\beta_2}{2\sqrt{3}} \\ -\frac{\beta_3}{2} - \frac{\beta_3}{2\sqrt{3}} & \;\; \frac{\beta_3}{2} - \frac{\beta_3}{2\sqrt{3}} & \frac{\beta_4
}{\sqrt{3}}
\end{array} \right ).
  \end{equation}
  In this section, we will focus on the diagonalization of the SM Yukawa couplings as given in Eq.~(\ref{eq:yu}), and save the discussion of the messenger Yukawa couplings for later in this work. 

It is straightforward to diagonalize Eq.~(\ref{eq:yu}) for arbitrary (real) $\beta_i$ via a standard biunitary transformation, in which 
  \begin{equation}
  U_{uL}^\dagger Y_u U_{uR} =Y_u^\text{diag},
  \end{equation}
with   
\begin{equation}
U_{uL}^\dagger Y_u Y_u^\dagger U_{uL}, \qquad U_{uR}^\dagger Y_u^\dagger Y_u U_{uR}. 
\end{equation}
 It is clear from the structure of Eq.~(\ref{eq:yu}) that the eigenvalues are not hierarchical for arbitary values of the $\beta_i$. Hence, specific relations among the $\beta_i$ are required for this scenario to be phenomenologically viable of this scenario.  Any such relations correspond to additional symmetry structures, together with the $\mathcal{S}_3$ Higgs-messenger symmetry.  
  
As discussed in \cite{Everett:2018wrn}, one possible solution that guarantees two zero mass eigenvalues and one nonzero mass eigenvalue is to enforce the following constraints:
\begin{equation}
\beta_1=1, \qquad \beta_2\beta_3=\beta_4.  
\label{eq:relations}
\end{equation}
The nonzero eigenvalue is then to be identified with the top quark Yukawa coupling, $y_t$. As discussed in \cite{Everett:2018wrn}, this requires the specific identification that $y_u=y_t/(\sqrt{2+\beta_2^2}\sqrt{2+\beta_3^2})$.
 
Furthermore, from Eq.~(\ref{eq:yu}) and Eq.~({\ref{eq:relations}), we see that one of the zero mass eigenvalues arises from the upper $2\times 2$ block of $Y_u$ and is controlled by $\beta_1\rightarrow 1$, while the other arises from the symmetry of the third column and row of $Y_u$ and is controlled by $\beta_2\beta_3\rightarrow \beta_4$.  Note that Eq.~(\ref{eq:relations}) includes the possibility that all $\beta_i=1$, for which there is the enhanced symmetry $\mathcal{S}_{3L}\times \mathcal{S}_{3R}$. This is the flavor ``democratic'' limit, which was studied in this context in \cite{Everett:2016meb}, and which has a long and extensive literature (see e.g.~\cite{harari,koide,tanimoto,fritzschplankl,Cvetic:1994sg,Fritzsch:1995dj,Xing:1996hi,Abel:1998wh,Mondragon:1998gy,Fritzsch:1998xs,Fritzsch:1999ee,Branco:2001hn, Rodejohann:2004qh,Gerard:2012ft,Fritzsch:2017tyf,Kaya:2018rsr,Ghosh:2018tzv}).   However, Eq.~(\ref{eq:relations}) also encompasses other possibilities. This includes the option that $\beta_4\gg \beta_{2,3}\gg \beta_1$, in which the term involving $\mathcal{S}_3$ singlet representations only in Eq.~(\ref{wu}) is dominant, which was explored in \cite{Everett:2018wrn}.

Given that there is a degenerate subspace corresponding to the two zero mass eigenvalues, the diagonalization matrices  $U_{uL}$ and $U_{uR}$ should generally involve linear combinations of the associated eigenvectors, with the linear combinations parametrized by a continuous parameter.  More precisely, the (unnormalized) eigenvector corresponding to the zero eigenvalue controlled by $\beta_1\rightarrow 1$ is given by $(1,-1,0)$, while the  (unnormalized) eigenvector corresponding to the other zero eigenvalue is given by $(-\beta_{3,2},-\beta_{3,2}, 1)$, with $\beta_{3,2}$ corresponding to the eigenvectors for $Y_u Y_u^\dagger$ and $Y_u^\dagger Y_u$, respectively. With this in mind, the diagonalization matrices $U_{uL}$ and $U_{uR}$ are given by
\begin{equation}
U_{uL}=\left (\begin{array}{ccc} \;\;\; \frac{\cos\tilde{\theta}}{\sqrt{2}}-\frac{\beta_3\sin\tilde{\theta}}{\sqrt{2}\sqrt{2+\beta_3^2}} & -\frac{\beta_3\cos\tilde{\theta}}{\sqrt{2}\sqrt{2+\beta_3^2}} -\frac{\sin\tilde{\theta}}{\sqrt{2}} & \frac{1}{\sqrt{2+\beta_3^2}} 
\\ -\frac{\cos\tilde{\theta}}{\sqrt{2}}- \frac{\beta_3\sin\tilde{\theta}}{\sqrt{2}\sqrt{2+\beta_3^2}} & -\frac{\beta_3\cos\tilde{\theta}}{\sqrt{2}\sqrt{2+\beta_3^2}} +\frac{\sin\tilde{\theta}}{\sqrt{2}} & \frac{1}{\sqrt{2+\beta_3^2}}
\\ \;\;\; \frac{\sqrt{2} \sin \tilde{\theta}}{\sqrt{2+\beta_3^2}}  &\frac{\sqrt{2} \cos\tilde{\theta}}{\sqrt{2+\beta_3^2}} & \frac{\beta_3}{\sqrt{2+\beta_3^2}} 
\end{array}\right )
\label{eq:uuL}
\end{equation}
\begin{equation}
U_{uR}=\left (\begin{array}{ccc} \;\;\; \frac{\cos\tilde{\theta}}{\sqrt{2}}-\frac{\beta_2\sin\tilde{\theta}}{\sqrt{2}\sqrt{2+\beta_2^2}} & -\frac{\beta_2\cos\tilde{\theta}}{\sqrt{2}\sqrt{2+\beta_2^2}} -\frac{\sin\tilde{\theta}}{\sqrt{2}} & \frac{1}{\sqrt{2+\beta_2^2}} 
\\ -\frac{\cos\tilde{\theta}}{\sqrt{2}}- \frac{\beta_2\sin\tilde{\theta}}{\sqrt{2}\sqrt{2+\beta_2^2}} & -\frac{\beta_2\cos\tilde{\theta}}{\sqrt{2}\sqrt{2+\beta_2^2}} +\frac{\sin\tilde{\theta}}{\sqrt{2}} & \frac{1}{\sqrt{2+\beta_2^2}}
\\ \;\;\; \frac{\sqrt{2} \sin \tilde{\theta}}{\sqrt{2+\beta_2^2}}  &\frac{\sqrt{2} \cos \tilde{\theta}}{\sqrt{2+\beta_2^2}} & \frac{\beta_2}{\sqrt{2+\beta_2^2}} 
\end{array}\right ),
\label{eq:uuR}
\end{equation}
in which we have written the linear combinations of degenerate eigenvectors in terms of the parameter $\tilde{\theta}$, with $0\leq \tilde{\theta}\leq \pi/2$.  In the case that $\tilde{\theta}=0$, the mass ordering is such that the eigenvalue controlled by $\beta_1$ would correspond to the first generation, and $U_{uL}$, $U_{uR}$ then reduce to the forms given in \cite{Everett:2018wrn}. In contrast, for $\tilde{\theta}=\pi/2$, it is the other eigenvalue that is to be identified with the first generation, and the corresponding $U_{uL}$, $U_{uR}$ have their first two columns interchanged compared to the forms given in \cite{Everett:2018wrn}. Of course, $\tilde{\theta}$ is an unphysical parameter in the degenerate (massless) limit, as studied in \cite{Everett:2018wrn}.  It is only when perturbations to this leading order structure are incorporated such that there are three distinct hierarchical mass eigenvalues that a specific value of $\tilde{\theta}$ is determined.  Indeed, a primary goal of this work is to explore such perturbations to see if viable quark masses and mixing can be obtained in this scenario.

To this end, we note that if identical structures are assumed within the down quark sector, such that the $U_{dL}$, $U_{dR}$ that satisfy $U_{dL}^\dagger Y_d U_{dR} = Y_d^{\rm diag}$ are given by Eqs.~(\ref{eq:uuL})--(\ref{eq:uuR}) with $\beta_{3,2}\rightarrow \beta_{3d,2d}$ and $\tilde{\theta} \rightarrow \tilde{\theta}_d$, the Cabibbo-Kobayashi-Maskawa matrix $U_{\rm CKM} = U_{uL}^\dagger U_{dL}$ takes the general form
\begin{equation}
U_{\rm CKM}=\left (\begin{array}{ccc} \;\;\; \cos\tilde{\theta} \cos\tilde{\theta}_d +\frac{(2+\beta_3\beta_{3d})\sin\tilde{\theta} \sin\tilde{\theta}_d}{\sqrt{2+\beta_3^2}\sqrt{2+\beta_{3d}^2}}
 & \frac{(2+\beta_3\beta_{3d})\cos\tilde{\theta}_d\sin\tilde{\theta}}{\sqrt{2+\beta_3^2}\sqrt{2+\beta_{3d}^2}}-\cos\tilde{\theta} \sin\tilde{\theta}_d
& -\frac{\sqrt{2}(\beta_3-\beta_{3d})\sin\tilde{\theta}}{\sqrt{2+\beta_3^2}\sqrt{2+\beta_{3d}^2}} 
\\ 
-\cos\tilde{\theta}_d \sin\tilde{\theta}+ \frac{(2+\beta_3\beta_{3d})\cos\tilde{\theta}\sin\tilde{\theta}_d}{\sqrt{2+\beta_3^2}\sqrt{2+\beta_{3d}^2}}
&  \frac{(2+\beta_3\beta_{3d})\cos\tilde{\theta}\cos\tilde{\theta}_d}{\sqrt{2+\beta_3^2}\sqrt{2+\beta_{3d}^2}}+\sin\tilde{\theta}\sin\tilde{\theta}_d 
& -\frac{\sqrt{2}(\beta_3-\beta_{3d})\cos\tilde{\theta}}{\sqrt{2+\beta_3^2}\sqrt{2+\beta_{3d}^2}} 
\\ \;\;\;  \frac{\sqrt{2}(\beta_3-\beta_{3d})\sin\tilde{\theta}_d}{\sqrt{2+\beta_3^2}\sqrt{2+\beta_{3d}^2}} &
\frac{\sqrt{2}(\beta_3-\beta_{3d})\cos\tilde{\theta}_d}{\sqrt{2+\beta_3^2}\sqrt{2+\beta_{3d}^2}} & \frac{2+\beta_3\beta_{3d}}{\sqrt{2+\beta_3^2}\sqrt{2+\beta_{3d}^2}}
\end{array}\right ).
\label{eq:CKMgen}
\end{equation}
There are several illuminating features of Eq.~(\ref{eq:CKMgen}).  First, Eq.(\ref{eq:CKMgen}) shows that the $2-3$ and $1-3$ mixing angles of $U_{\rm CKM}$ both depend linearly on the quantity $\beta_3-\beta_{3d}$. In contrast, the Cabibbo ($1-2$) mixing angle $\lambda$ is largely independent of this factor, and instead depends primarily on the difference between $\tilde{\theta}$ and $\tilde{\theta}_d$. In the case that $\tilde{\theta}$ takes intermediate values such that $\sin\tilde{\theta}\sim \cos\tilde{\theta}$, it is necessary to take $\beta_{3d}\rightarrow \beta_3+O(\lambda^3)$. This a very delicate balance that is needed between the up and down quark sectors, and ensuring that this condition is satisfied certainly requires additional model-building input. In this case, further corrections are required to fill in the needed value of the $2-3$ CKM mixing angle. Second, it is possible to envision a scenario in which $\tilde{\theta}\rightarrow O(\lambda)$, such that we can take the still-stringent but slightly milder condition that $\beta_{3d}\rightarrow \beta_3+O(\lambda^2)$.  Indeed, in the limit that $\beta_{3d}\rightarrow \beta_3$, Eq.~({\ref{eq:CKMgen}) simplifies to the following form:
\begin{equation}
U_{\rm CKM} = \left ( \begin{array}{ccc} \cos(\tilde{\theta}-\tilde{\theta}_d) & \sin (\tilde{\theta}-\tilde{\theta}_d) & 0 \\ -\sin (\tilde{\theta}-\tilde{\theta}_d) & \cos (\tilde{\theta}-\tilde{\theta}_d) & 0 \\ 0 & 0 & 1 \end{array} \right ).
\label{eq:ckmsimplified}
\end{equation}
Clearly in this case we must also have $\tilde{\theta}_d\rightarrow O(\lambda)$ and $\tilde{\theta}-\tilde{\theta}_d\sim O(\lambda)$. This is also a delicate balance between the up and down quark sectors, and further model-building structure must be incorporated to generate such relations dynamically rather than achieving them via fine-tuning.

While it might at first seem plausible that perturbations to Eq.~(\ref{eq:relations}) could yield a phenomenologically acceptable CKM matrix, we can see right away that this is impossible.  The reason is that Eq.~({\ref{eq:yu}) is exactly diagonalizable for arbitrary $\beta_i$. Hence,  once the $\beta_i$ no longer satisfy Eq.~(\ref{eq:relations}), the hierarchy of the eigenvalues is immediately fixed (up to the possible but uninteresting case of degenerate but nonzero masses) such that either we have $\tilde{\theta}=0$ or $\tilde{\theta}=\pi/2$ (and analogous results for $\tilde{\theta}_d$), with no small corrections to either of these cases. As a result, the prediction for the Cabibbo angle at the renormalizable level is either vanishingly small if $\tilde{\theta}=\tilde{\theta}_d$, or $O(1)$ if $\tilde{\theta}-\tilde{\theta}_d\sim \pi/2$, which are both phenomenologically unacceptable. This leads us to consider nonrenormalizable operators that can contribute to the SM fermion masses, as discussed in the next section.

%\noindent {\bf SM and Messenger Yukawa couplings: nonrenormalizable operators}. 

\section{Fermion Masses: Nonrenormalizable Couplings}

As discussed in the previous section, the Yukawa couplings at the renormalizable level do not give rise to a phenomenologically acceptable CKM matrix.  Hence, we now explore the possibility that the renormalizable couplings listed in Eq.~(\ref{UpYukawas}) are supplemented by couplings of the Higgs-messenger fields to the matter fields that are induced at the nonrenormalizable level.  

Given the quantum numbers of the fields of the theory as given in Table~\ref{tab:12}, it is clear that this requires augmenting the theory to include a flavon sector that consists of additional superfields that are assumed to have vacuum expectation values in their scalar components (but no associated $F$ terms). Furthermore, it is clear that the flavon sector must include fields with nontrivial $\mathcal{S}_3$ quantum numbers, which then can easily resemble the corresponding $\mathcal{H}^{(2)}_{u,d}$ fields.  Quite generally, with the introduction of such flavon fields, additional model-building constraints are required to ensure, for example, that such flavons do not couple directly to the $X_{H,T}$ fields of the theory, for example.  Our purpose in this work is to not to provide a comprehensive analysis that includes the details of the flavon sector dynamics, but rather to provide an explicit working example of an nonrenormalizable operator that can satisfy the requirements of the previous section for generating a viable Cabibbo mixing angle.  

The working example we construct is as follows. Let us consider the following coupling:
\begin{equation}
W^{(u)}_{NR}=\frac{\epsilon}{\Lambda'} \Big(Q_{\mathbf 2} \phi_{\mathbf 2} \Big)\Big(\mathcal{H}_u^{(2)} \bar u_{\mathbf 2}\Big),
\label{nonrnoperator}
\end{equation}
in which $\phi_{\mathbf 2}$ is a flavon in the ${\mathbf 2}$ representation of $\mathcal{S}_3$. Here $\epsilon$ is a dimensionless parameter, and $\Lambda^\prime$ is the scale of the new physics that is responsible for generating this operator. Through some dynamics (that as stated we will leave unspecified in this work), $\phi_{\mathbf 2}$ acquires a vacuum expectation value in its scalar component, but as previously just discussed, not its $F$-component, so it does not participate in the mediation of supersymmetry breaking. We parametrize this field's vacuum expectation value as
\begin{equation}
\langle \phi_{\mathbf 2}\rangle =
v\left(\begin{array}{cc} 
\cos\theta \\
\sin\theta
\end{array}
\right),
\end{equation}
in which $v$ is a dimensionful parameter, and the dimensionless parameter $\theta$ has been introduced (and we will shortly see its identification with the parameter $\theta$  as given in the previous section). After this flavon acquires a vev, the strength of its nonrenormalizable coupling in Eq.~(\ref{nonrnoperator}) is given by $\beta_\epsilon\equiv v\epsilon/\Lambda'$. We then obtain an additional contribution to the SM up quark Yukawa matrix:
\begin{equation}
Y^{NR}_u=\frac{\beta_\epsilon}{\sqrt{3}}\left(
\begin{array}{ccc}
 \sin \theta  & \sin \theta  & 0 \\
 \cos \theta  & \cos \theta  & 0 \\
 0 & 0 & 0 \\
\end{array}
\right),
\label{upyukcorr}
\end{equation}
as well as contributions to the messenger Yukawa couplings,
\begin{align}
Y'^{NR}_{u1}&=\frac{\beta_\epsilon}{2}\left(
\begin{array}{ccc}
 \left(1-\frac{1}{\sqrt{3}}\right) \sin \theta  & -\left(1+\frac{1}{\sqrt{3}}\right) \sin \theta  & 0 \\
 \left(1-\frac{1}{\sqrt{3}}\right) \cos \theta  & - \left(1+\frac{1}{\sqrt{3}}\right) \cos \theta  & 0 \\
 0 & 0 & 0 \\
\end{array}
\right)\nonumber\\
Y'^{NR}_{u2}&=\frac{\beta_\epsilon}{2}\left(
\begin{array}{ccc}
 - \left(1+\frac{1}{\sqrt{3}}\right) \sin \theta  & \left(1-\frac{1}{\sqrt{3}}\right) \sin \theta  & 0 \\
 - \left(1+\frac{1}{\sqrt{3}}\right) \cos \theta  & \left(1-\frac{1}{\sqrt{3}}\right) \cos \theta  & 0 \\
 0 & 0 & 0 \\
\end{array}
\right),
\label{messyukscorr}
\end{align}
and we assume there are analogous relations for the down quark and charged lepton sectors.
The task at hand is once again to diagonalize the SM Yukawa couplings, which now take the form $Y_u \rightarrow Y_u+Y^{NR}_u$.  Again, in this section we will focus on the SM fermion masses, and defer the discussion of the associated messenger Yukawa couplings to the next section.

Here we will focus our attention on the case in which we retain the relations of Eq.~(\ref{eq:relations}) for the renormalizable couplings, such that $\beta_1=1$ and $ \beta_2\beta_3=\beta_4$.  The SM up quark Yukawa matrix then takes the form
\begin{equation}
Y_u\rightarrow \frac{1}{\sqrt{3}}\left(
\begin{array}{ccc}
y_u+\beta_\epsilon \sin\theta & y_u+\beta_\epsilon\sin\theta & y_u \beta_2 \\ 
y_u+\beta_\epsilon \cos\theta & y_u+\beta_\epsilon \cos\theta & y_u \beta_2 \\
y_u \beta_3 & y_u \beta_3 & y_u \beta_2 \beta_3
\end{array}
\right).
\label{eq:fullyu}
\end{equation}
Diagonalizing this matrix in the usual manner, the eigenvalues are easily shown to be nondegenerate. As we will see, one eigenvalue remains massless, the second has mass of order $\beta_\epsilon \ll 1$, and the third is to be identified with the top quark Yukawa coupling $y_t$.   

While it is straightforward to obtain the diagonalization matrices for arbitrary values of the parameters $\beta_{2,3}$ and $\beta_\epsilon$, here we focus on leading order effects in $\beta_\epsilon$. We also focus here on the limit 
investigated in \cite{Everett:2018wrn}, wherein $\beta_{2,3}$ are taken to be very large while $y_u$ is taken such that $y_t$ remains constant. This is done not only for simplicity, but also because deviations from that limit generically result in flavor off-diagonal couplings in the messenger sector, which require more detailed analysis \footnote{A notable exception is the ``democratic'' limit in which the $\beta_i$ couplings of the renormalizable sector are all equal to 1; we defer a detailed discussion of this case to a future study.}.

In this paradigm, the SM up quark Yukawa becomes
\begin{equation}
Y_u=\left(
\begin{array}{ccc}
 \frac{\beta_\epsilon  \sin \theta }{\sqrt{3}} & \frac{\beta_\epsilon  \sin \theta }{\sqrt{3}} & 0 \\
 \frac{\beta_\epsilon  \cos \theta }{\sqrt{3}} & \frac{\beta_\epsilon  \cos \theta }{\sqrt{3}} & 0 \\
 0 & 0 & y_t \\
\end{array}
\right),
\label{yunrlim}
\end{equation}
while the messenger Yukawas take the form
\begin{align}
Y'_{u1}=&\left(
\begin{array}{ccc}
 \beta_\epsilon\left(\frac{1}{2}-\frac{1}{2\sqrt{3}}\right) \sin \theta  & -\beta_\epsilon \left(\frac{1}{2}+\frac{1}{2\sqrt{3}}\right) \sin \theta  & 0 \\
 \beta_\epsilon\left(\frac{1}{2}-\frac{1}{2\sqrt{3}}\right) \cos \theta  & -\beta_\epsilon \left(\frac{1}{2}+\frac{1}{2\sqrt{3}}\right) \cos \theta  & 0 \\
 0 & 0 & y_t  \\
\end{array}
\right)\nonumber\\
Y'_{u2}=&\left(
\begin{array}{ccc}
 -\beta_\epsilon\left(\frac{1}{2}+\frac{1}{2\sqrt{3}}\right) \sin \theta  & \beta_\epsilon\left(\frac{1}{2}-\frac{1}{2\sqrt{3}}\right) \sin \theta  & 0 \\
 -\beta_\epsilon\left(\frac{1}{2}+\frac{1}{2\sqrt{3}}\right) \cos \theta  & \beta_\epsilon\left(\frac{1}{2}-\frac{1}{2\sqrt{3}}\right) \cos \theta  & 0 \\
 0 & 0 & y_t \\
\end{array}
\right).
\label{ymesslim}
\end{align}
Upon first inspection, it appears that the Yukawa matrices are dependent on the direction of the flavon vacuum expectation value, and as such one might expect the eigenvalues of $Y_u$ to also carry this dependence. However, this is not the case, as we will soon see.

Following the standard procedure of rotating the SM up quark Yukawa in Eq. (\ref{yunrlim}) into the diagonal quark mass basis using a biunitary transformation, the diagonalization matrices $U_{uL}$ and $U_{uR}$ are found to take the simple forms
\begin{align}
U_{uL}&=\left(
\begin{array}{ccc}
 -\cos\theta & \sin\theta & 0 \\
 \sin\theta & \cos\theta & 0 \\
 0 & 0 & 1 \\
\end{array}
\right) \qquad
%\\
U_{uR}=\left(
\begin{array}{ccc}
 -\frac{1}{\sqrt{2}} & \frac{1}{\sqrt{2}} & 0 \\
 \frac{1}{\sqrt{2}} & \frac{1}{\sqrt{2}} & 0 \\
 0 & 0 & 1 \\
\end{array}
\right).
\end{align} 
In this basis, the SM up quark Yukawa matrix is
\begin{equation}
Y_u={\rm Diag}\left(0,\sqrt{\frac{2}{3}}\beta_\epsilon,y_t\right).
\label{yulimdiag}
\end{equation}
We now see that the dependence on the direction of the vacuum expectation value of the flavon field drops out in the Yukawa matrices, but is now carried by the unitary matrices $U_{u:}$ and $U_{uR}$. It is immediately clear, however, that $\theta$ enters into the flavor structure of this model. Explicitly, assuming a corresponding structure in the down quarks, ($\theta\to\theta_d$, $y_u\to y_d$, and $\beta_i\to\beta_{di}$), we obtain a CKM matrix of the form 
\begin{equation}
U_{\rm CKM}=U_{uL}^{\dagger}U_{dL}=\left(
\begin{array}{ccc}
 \cos (\theta-\theta_d) & \sin (\theta-\theta_d) & 0 \\
 -\sin (\theta-\theta_d) & \cos (\theta-\theta_d) & 0 \\
 0 & 0 & 1 \\
\end{array}
\right).
\label{ckmcase2}
\end{equation}
We note that the structure of the CKM as given in Eq. (\ref{ckmcase2}) is unambiguous, as the quark masses are non-degenerate (as seen in Eq. (\ref{yulimdiag})), and it describes mixing between the first and second generations. This form explicitly allows for the generation of appropriately Cabibbo-sized mixing between the first and second families if $\sin(\theta-\theta_d)\simeq \lambda$, as anticipated from the general discussion of the last section. Indeed, upon a comparison of Eq.~(\ref{ckmcase2}) with the renormalizable level structure of Eq.~(\ref{eq:CKMgen}), we see that we have obtained a viable CKM matrix to leading order in the Cabibbo angle $\lambda$, and that the parameters $\tilde{\theta}$ and $\tilde{\theta}_d$ of the previous section can be identified with the quantities $\theta$ and $\theta_d$ of this section, which parametrize the vacuum expectation values of the flavon fields in the up and down quark sectors (compare Eq.~(\ref{eq:ckmsimplified}) and Eq.~(\ref{ckmcase2})).  Furthermore, we note that as we have explicitly taken the limit that $\beta_{2,3}\gg 1$ and $\beta_{2d,3d}\gg 1$, at this order no $1-3$ or $2-3$ CKM mixing is generated.  To summarize, this operator has indeed led to a working example of lifting the mass degeneracy of the couplings of the renormalizable sector in the case that $\beta_1=1$, $\beta_4=\beta_2\beta_3$ (and analogously for the down quark sector), in such a way that a Cabibbo mixing angle of the appropriate size can be generated.

\section{Messenger Yukawa Couplings and Superpartner Mass Spectra}

In this section, we turn our attention to the messenger Yukawa couplings and resulting mass spectra of the MSSM superpartners.  Here we will confine our attention to the large $\beta_{2,3}$ regime, for which the structure of the resulting soft terms is particularly simple, and is flavor diagonal.  We defer a more comprehensive analysis of general $\beta_{2,3}$ that satisfy Eq.~(\ref{eq:relations}) for a future study.\\

\noindent{\bf Messenger Yukawa couplings and soft supersymmetry breaking terms}. We begin by writing the messenger Yukawa couplings in the diagonal SM fermion mass basis.  For the up quark sector, it is straightforward to determine that starting from Eq.~(\ref{messyukscorr}), the messenger Yukawas in the diagonal quark mass basis, in the limit that $\beta_{2,3}\gg 1$, are given by
\begin{equation}
Y'_{u1}=\left(
\begin{array}{ccc}
0&0&0\\
-\frac{\beta_\epsilon}{\sqrt{2}}&-\frac{\beta_\epsilon}{\sqrt{6}}&0 \\ 
0&0&y_t
\end{array} \right )
\qquad 
Y'_{u2}=\left(
\begin{array}{ccc}
0&0&0\\
\frac{\beta_\epsilon}{\sqrt{2}}&-\frac{\beta_\epsilon}{\sqrt{6}}&0 
\\ 0&0&y_t
\end{array}\right).
\label{ymlimdiag}
\end{equation}
With these simple forms of the up quark and messenger Yukawa matrices as given in Eqs. (\ref{yulimdiag}) and (\ref{ymlimdiag}), the corrections to the soft supersymmetry breaking terms are easily calculated. The methods for doing so are standard in the literature, (see e.g. \cite{Abdullah:2012tq}, \cite{Evans:2013kxa}, \cite{Jelinski:2015voa}), and are summarized for these classes of models in \cite{Everett:2016meb}. 

As before, we assume that the doublet and triplet messengers are determined by the same value of   $\Lambda=F_{2,3}/M_{Mess}\approx F/M,$ and that the down quark and charged lepton sectors are analogous to the up quark sector. As a first step in exploring the phenomenology of this scenario, and to examine in detail the effects of the nonrenormalizable operator of Eq.~(\ref{nonrnoperator}), for simplicity we assume a single $\beta_\epsilon$ parameter for each of the sectors, and allow it to vary (while keeping $\beta_\epsilon \ll 1$).  The soft terms include the usual gauge-mediated contributions (not shown for simplicity), as well as corrections due to the messenger Yukawa couplings.  

The nonvanishing corrections to the soft terms from the messenger Yukawas are presented below (here the relevant factors of $\Lambda/(4\pi)^2$ are suppressed for notational convenience):
\begin{align}
\left(\delta m^2_{\tilde{Q}}\right)_{22}&=\left(-2 y_b^2-2 y_t^2-\frac{2y_\tau^2}{3}-\frac{16 g_1^2}{9}-8 g_2^2-\frac{128 g_3^2}{9}\right)\beta_\epsilon ^2 \nonumber \\
\left(\delta m^2_{\tilde{Q}}\right)_{33}&= 36 y_b^4+8 y_t^2 y_b^2+8y_\tau^2 y_b^2-\frac{14 g_1^2 y_b^2}{15}-6 g_2^2
y_b^2-\frac{32 g_3^2 y_b^2}{3}+36 y_t^4-\frac{26 g_1^2 y_t^2}{15}-6 g_2^2 y_t^2-\frac{32g_3^2 y_t^2}{3}\nonumber\\
&+\left(-\frac{8 y_b^2}{3}-2 y_t^2\right) \beta_\epsilon ^2 \nonumber \\
\left(\delta m^2_{\tilde{u}}\right)_{11}&=\left(-\frac{26 g_1^2}{15}-6 g_2^2-\frac{32 g_3^2}{3}\right) \beta_\epsilon ^2, \qquad
\left(\delta m^2_{\tilde{u}}\right)_{22}= \left(-4 y_t^2-\frac{26 g_1^2}{45}-2 g_2^2-\frac{32 g_3^2}{9}\right) \beta_\epsilon ^2  \nonumber \\
\left(\delta m^2_{\tilde{u}}\right)_{33}&=72 y_t^4+8 y_b^2 y_t^2-\frac{52 g_1^2 y_t^2}{15}-12 g_2^2
   y_t^2-\frac{64 g_3^2 y_t^2}{3} -4 \beta_\epsilon ^2 y_t^2\nonumber \\
\left(\delta m^2_{\tilde{d}}\right)_{11}&=\left(4y_\tau^2-\frac{14 g_1^2}{15}-6 g_2^2-\frac{32 g_3^2}{3}\right) \beta_\epsilon ^2, \;\;\;
\left(\delta m^2_{\tilde{d}}\right)_{22}=\left(-4 y_b^2-\frac{14 g_1^2}{45}-2 g_2^2-\frac{32 g_3^2}{9}\right) \beta_\epsilon ^2  \nonumber \\
\left(\delta m^2_{\tilde{d}}\right)_{33}&=72 y_b^4+8 y_t^2 y_b^2+24y_\tau^2 y_b^2-\frac{28 g_1^2 y_b^2}{15}-12 g_2^2
   y_b^2-\frac{64 g_3^2 y_b^2}{3} \nonumber \\
\left(\delta m^2_{\tilde{L}}\right)_{22}&= \left(-2 y_b^2-\frac{2y_\tau^2}{3}-\frac{12 g_1^2}{5}-4 g_2^2\right) \beta_\epsilon ^2 \nonumber \\
\left(\delta m^2_{\tilde{L}}\right)_{33}&=20y_\tau^4+24 y_b^2y_\tau^2-\frac{18 g_1^2y_\tau^2}{5}-6
   g_2^2y_\tau^2-\frac{8 \beta_\epsilon ^2y_\tau^2}{3} \nonumber \\
\left(\delta m^2_{\tilde{e}}\right)_{11}&=\left(-\frac{18 g_1^2}{5}-6 g_2^2\right) \beta_\epsilon ^2 \qquad 
%\\
\left(\delta m^2_{\tilde{e}}\right)_{22}=\left(-4 y_b^2-\frac{4y_\tau^2}{3}-\frac{6 g_1^2}{5}-2 g_2^2\right) \beta_\epsilon ^2  \nonumber \\
\left(\delta m^2_{\tilde{e}}\right)_{33}&= 40y_\tau^4+48 y_b^2y_\tau^2-\frac{36 g_1^2y_\tau^2}{5}-12
   g_2^2y_\tau^2-\frac{16 \beta_\epsilon ^2y_\tau^2}{3}  \nonumber \\
\delta m^2_{\tilde{H}_u}&=-6 y_t^2 (y_b^2 + 3 y_t^2), \qquad 
%\\
\delta m^2_{\tilde{H}_d}=-6 (3 y_b^4 + y_b^2 y_t^2 + 3 y_\tau^4) \nonumber \\
\left(\tilde{A}_{u}\right)_{33}&=-2 y_t \left(y_b^2+3 y_t^2\right), \qquad
%\\
\left(\tilde{A}_{d}\right)_{33}= -2 y_b \left(3 y_b^2+y_t^2\right), \qquad
\left(\tilde{A}_{e}\right)_{33}= -6 y_\tau^3.
 \label{softterms}
\end{align}
We see that there is no introduction of off-diagonal flavor-violating couplings at leading order in this limiting case in which for the up, down, and charged lepton sectors,  Eq.~(\ref{eq:relations}) is satisfied and the relevant $\beta_{2,3}$ are taken to be very large while keeping the third generation SM fermion masses fixed. Furthermore, the corrections to the first two generations arise at order $\beta_\epsilon^2$.\\

\noindent {\bf Superpartner mass spectra}. We now explore the phenomenology of this scenario that arises from the soft terms as given in Eqs. (\ref{softterms}). As is {\it de rigueur}, our model parameters are $M_{\rm Mess},$ $\Lambda,$ $\tan\beta,$ and the sign of $\mu,$ where we have replaced $\mu$ and $b$ by $\tan\beta$, ${\rm sgn}(\mu)$ and the $Z$ boson mass.  We set ${\rm sgn}(\mu)=1.$ The renormalization group equations are run using SoftSUSY 4.1.4 \cite{Allanach:2001kg}.

In previous work \cite{Everett:2018wrn}, we explored the behavior of the superpartner mass spectra for the renormalizable sector Yukawa couplings in the large $\beta_i$ limit, focusing on the dependence of the spectra on $\tan\beta$ and the messenger scale. For continuity, as well as a check on the phenomenological consistency of the nonrenormalizable operator introduced in Eq. (\ref{nonrnoperator}), we begin with the example spectra as shown in Figure \ref{fig:CompareOff}. The left-hand side of Figure \ref{fig:CompareOff} shows results for the model studied in \cite{Everett:2018wrn}. The messenger scale is $M_{\text{Mess}}= 1\times10^{12}$ GeV and $\tan\beta=10$. The value of $\Lambda$ is set such that $m_h\simeq 125$ GeV. The right-hand side of Figure \ref{fig:CompareOff} displays the Higgs and superpartner mass spectra that arise from the soft supersymmetry breaking terms as given in Eq. (\ref{softterms}), but with $\beta_\epsilon=0$. The spectra are in agreement, as expected.

In the $\beta_\epsilon \rightarrow 0$ limit, the heavy Higgs particles are between $5-6$ TeV, along with the gluino at around $5$ TeV. The squark masses fall into two general categories, one significantly heavier than the other. The heavier squarks are the left-handed sdown, sup and scharm squarks, as well as the both scharms and the heavier of the two stops. Their masses are close to the heavier charginos and neutralinos. The lower group is comprised of the right-handed sdown, sup and scharm squarks, as well as both sbottoms, and the lighter stop, whose masses are closer to the gluino. The next-to-lightest supersymmetric particle (NSLP) in this scenario is a bino-like neutralino.

\begin{figure}
\begin{subfigure}
   \centering
   \includegraphics[width = 0.4\textwidth]{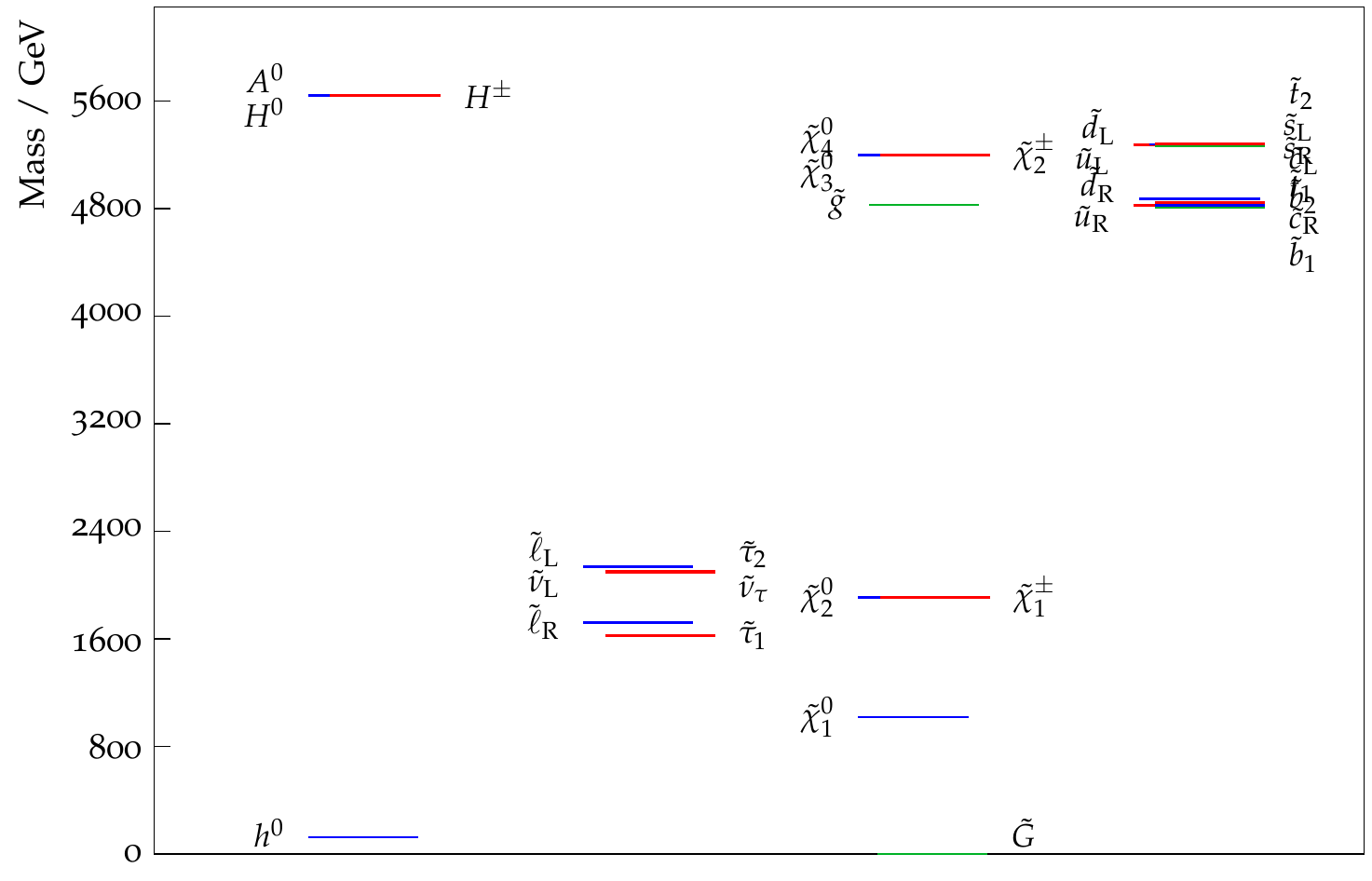} 
   \end{subfigure}
   \begin{subfigure}
   \centering
   \includegraphics[width = 0.4\textwidth]{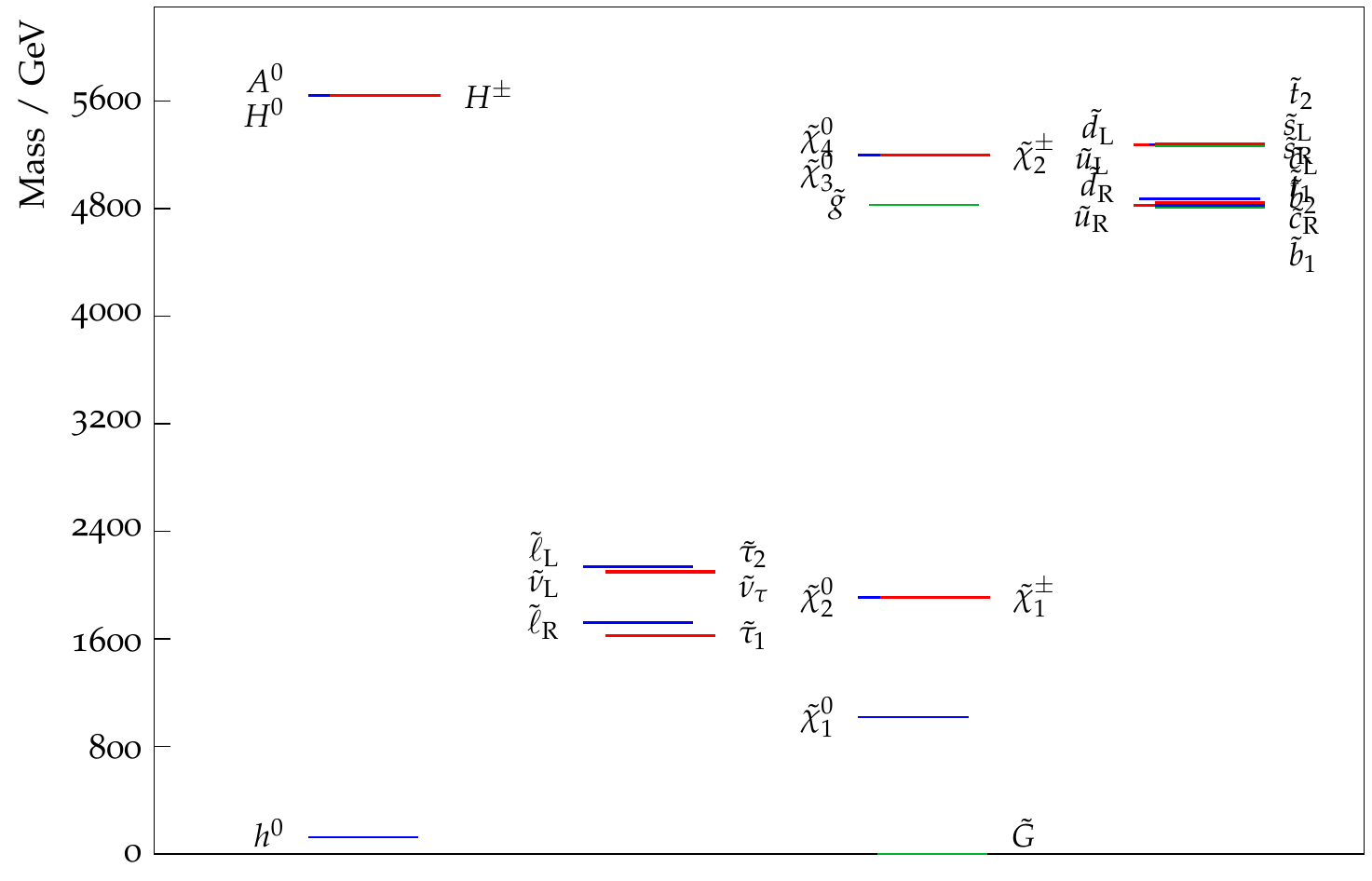}
   \end{subfigure}
   \centering
      \caption{The mass spectra for $M_{\text{Mess}}= 1\times10^{12}$ GeV and $\tan\beta=10$ for the scenario explored in \cite{Everett:2018wrn} and for the case explored here with $\beta_\epsilon=0$ (right). In each case, $\Lambda$ is fixed by the Higgs mass constraint. As expected, the two cases are in agreement.}
   \label{fig:CompareOff}
\end{figure}

Let us now include the effects of the nonrenormalizable operator given in Eq.~(\ref{nonrnoperator}), such that $\beta_\epsilon$ is now nonzero. Here we note that as $\beta_\epsilon$ is connected with the masses of the charged SM fermions of the second generation, this quantity is expected to take small values.  As explicit examples, the resulting spectra for small values of the coupling strengths $\beta_\epsilon$ are given in Figure \ref{fig:HighMessLowEps}. The left-hand side of Figure \ref{fig:HighMessLowEps} has taken $\beta_\epsilon=.01$, and the right-hand side has $\beta_\epsilon=.02$. The spectra follow the same general pattern as seen in Figure \ref{fig:CompareOff}, but with minor changes to the splitting of the superpartner masses. We see that the masses of the right-handed down and up squarks are pushed down, as well as the lightest left-handed charged slepton and sneutrino. The masses of the heavy Higgses are almost entirely unaffected, as are the masses of the gauginos.

\begin{figure}
\begin{subfigure}
   \centering
   \includegraphics[width = 0.4\textwidth]{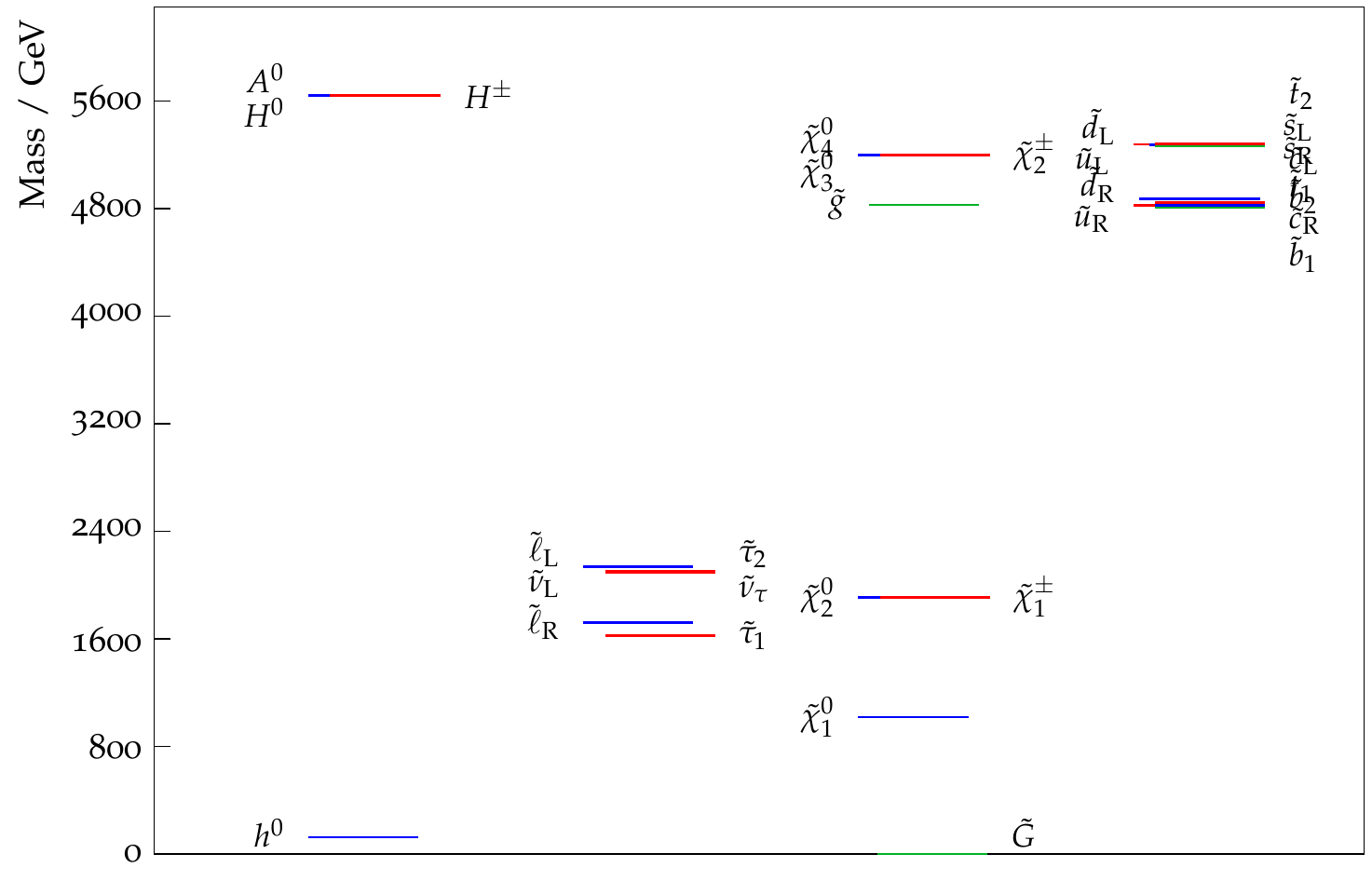}
   \end{subfigure}
 \begin{subfigure}
  \centering
   \includegraphics[width = 0.4\textwidth]{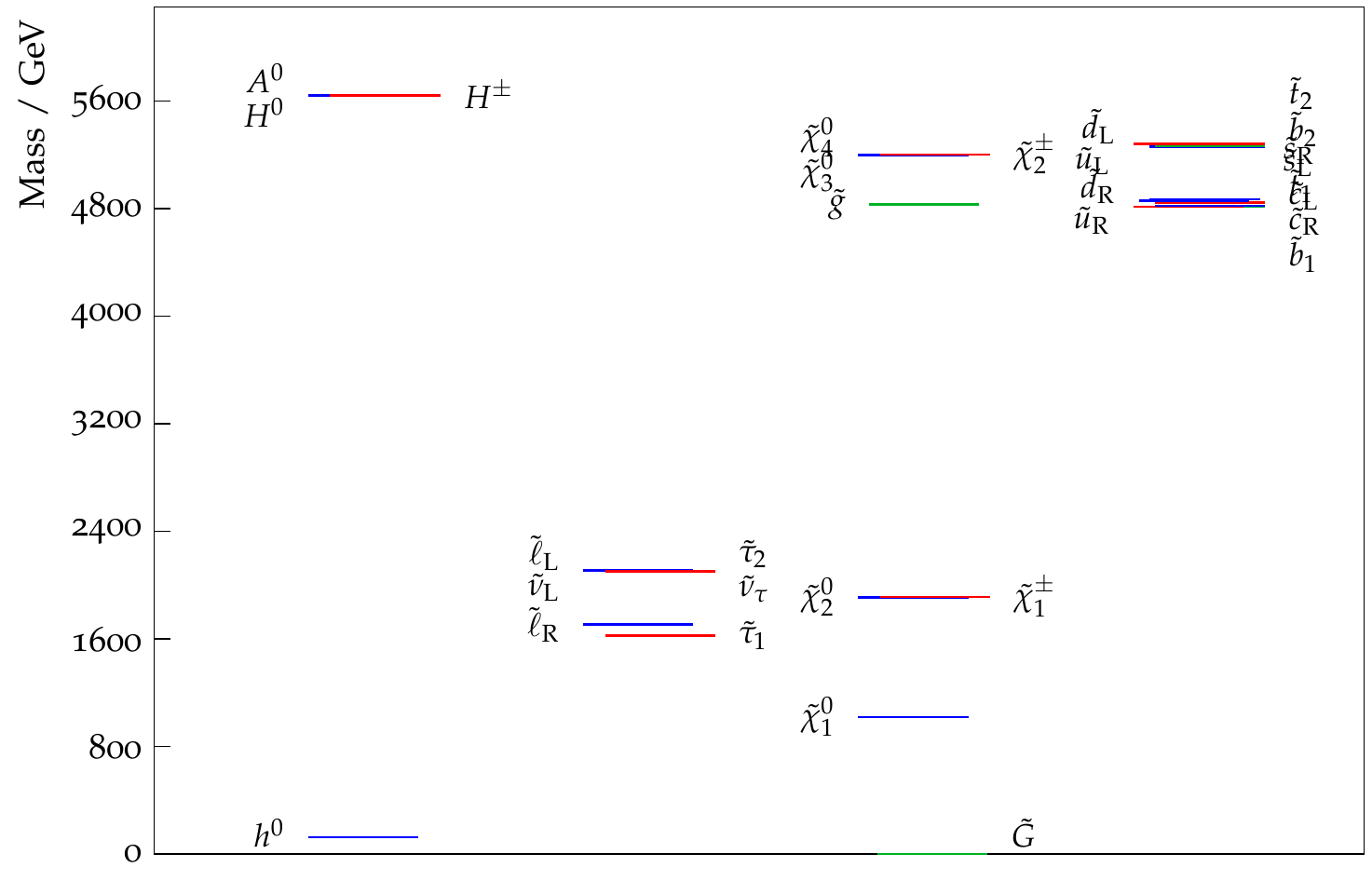}
   \end{subfigure}
   \centering
      \caption{Mass spectra for $M_{\text{Mess}}= 1\times10^{12}$ GeV (both sides), $\tan\beta=10$ and $\beta_\epsilon=0.01$ (left) and $\beta_\epsilon=0.05$ (right).  In each case, $\Lambda$ is fixed by the Higgs mass constraint.}
   \label{fig:HighMessLowEps}
\end{figure}

\begin{figure}
\begin{subfigure}
   \centering
   \includegraphics[width = 0.4\textwidth]{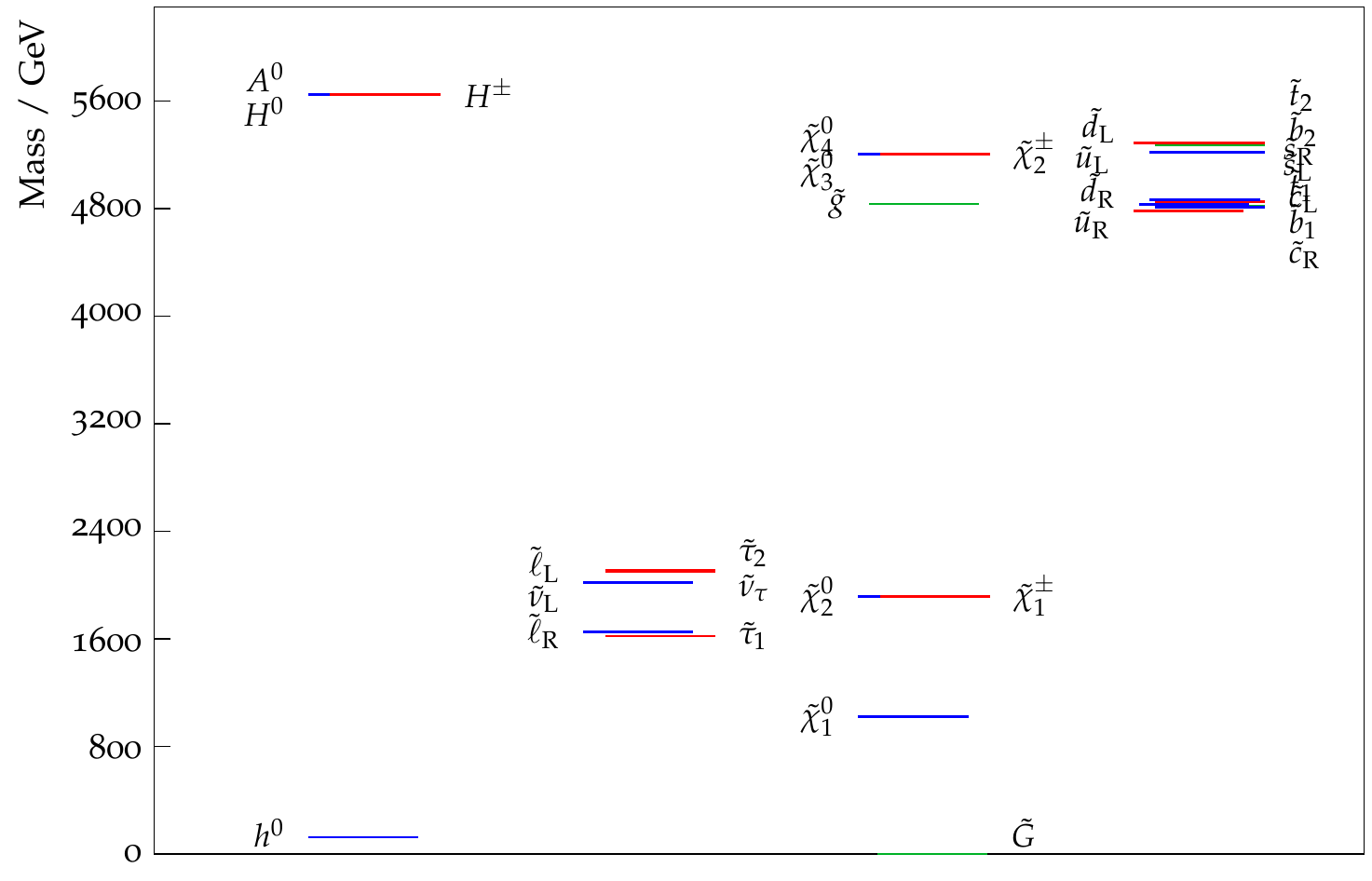}
   \end{subfigure}
 \begin{subfigure}
  \centering
   \includegraphics[width = 0.4\textwidth]{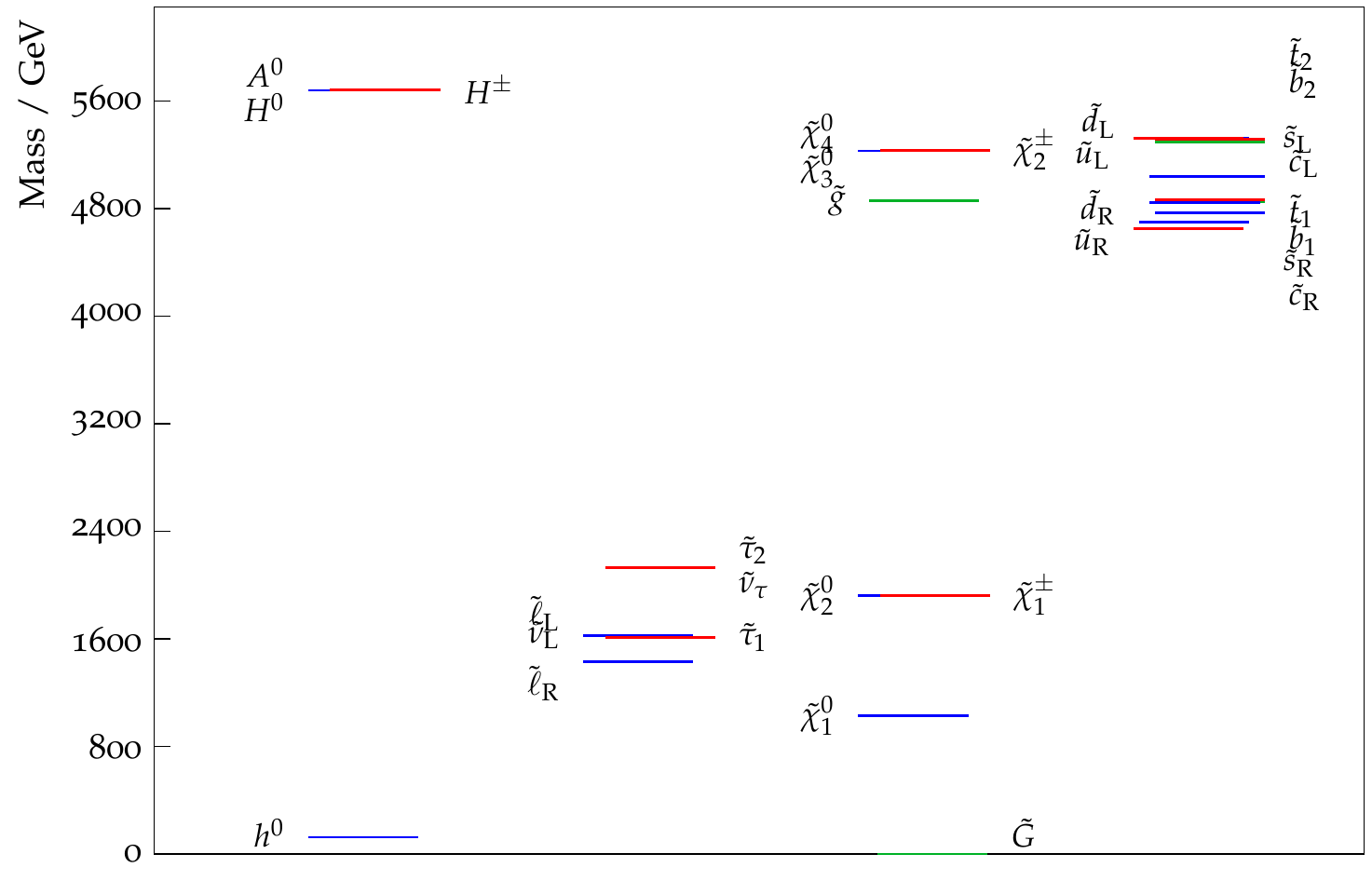}
   \end{subfigure}
   \centering
    \caption{Mass spectra for $M_{\text{Mess}}= 1\times10^{12}$ GeV (both sides), $\tan\beta=10$ and $\beta_\epsilon=0.1$ (left) and $\beta_\epsilon=0.2$ (right).  In each case, $\Lambda$ is fixed by the Higgs mass constraint.}
   \label{fig:HighMessBigEps}
\end{figure}

It is illustrative to consider what occurs for larger values of $\beta_\epsilon$, for comparative purposes (note that significant values of $\beta_\epsilon$ are inconsistent with SM charged fermion mass predictions). We find that the  pattern described above continues for such larger values of $\beta_\epsilon,$ as shown in Figure \ref{fig:HighMessBigEps}. On the left-hand side, for $\beta_\epsilon=0.1$, we see that the first two families of charged sleptons and sneutrinos are now lighter than one of the staus, with the other stau being the lightest slepton. Additionally, we see that the tight groupings of the squarks into two bands, as seen in Figure \ref{fig:CompareOff} are splitting with the right-handed sup becoming the lightest colored superpartner. On the right-hand side, which has $\beta_\epsilon=0.2,$ the lightest right-handed charged slepton is now lighter than all third generation sleptons. Furthermore, the squarks continue to display larger mass splittings, with the mass splittings within the original two groupings that appeared for smaller $\beta_\epsilon$ clearly demonstrated.

\begin{figure}
\begin{subfigure}
   \centering
   \includegraphics[width = 0.4\textwidth]{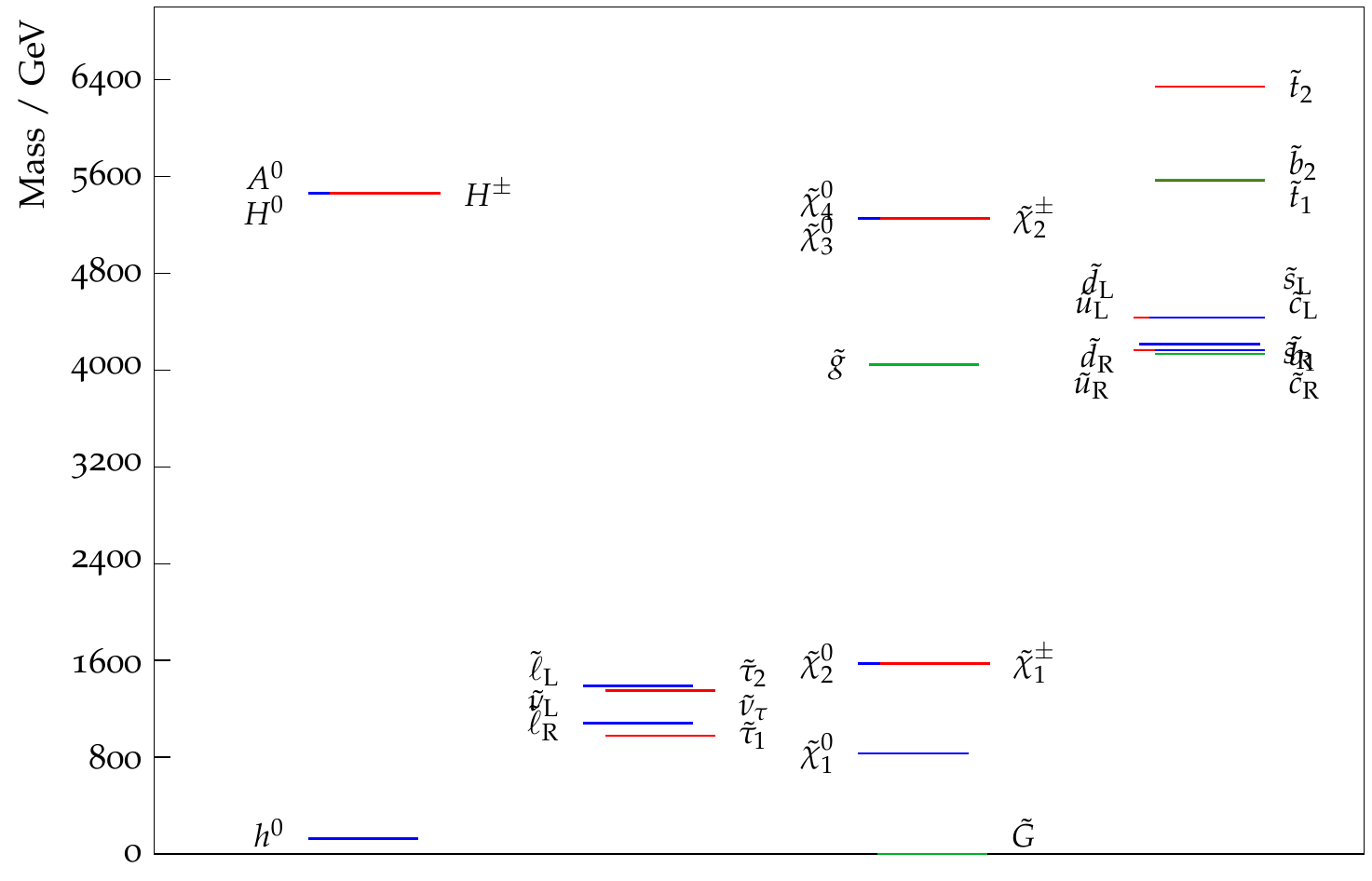}
   \end{subfigure}
 \begin{subfigure}
  \centering
   \includegraphics[width = 0.4\textwidth]{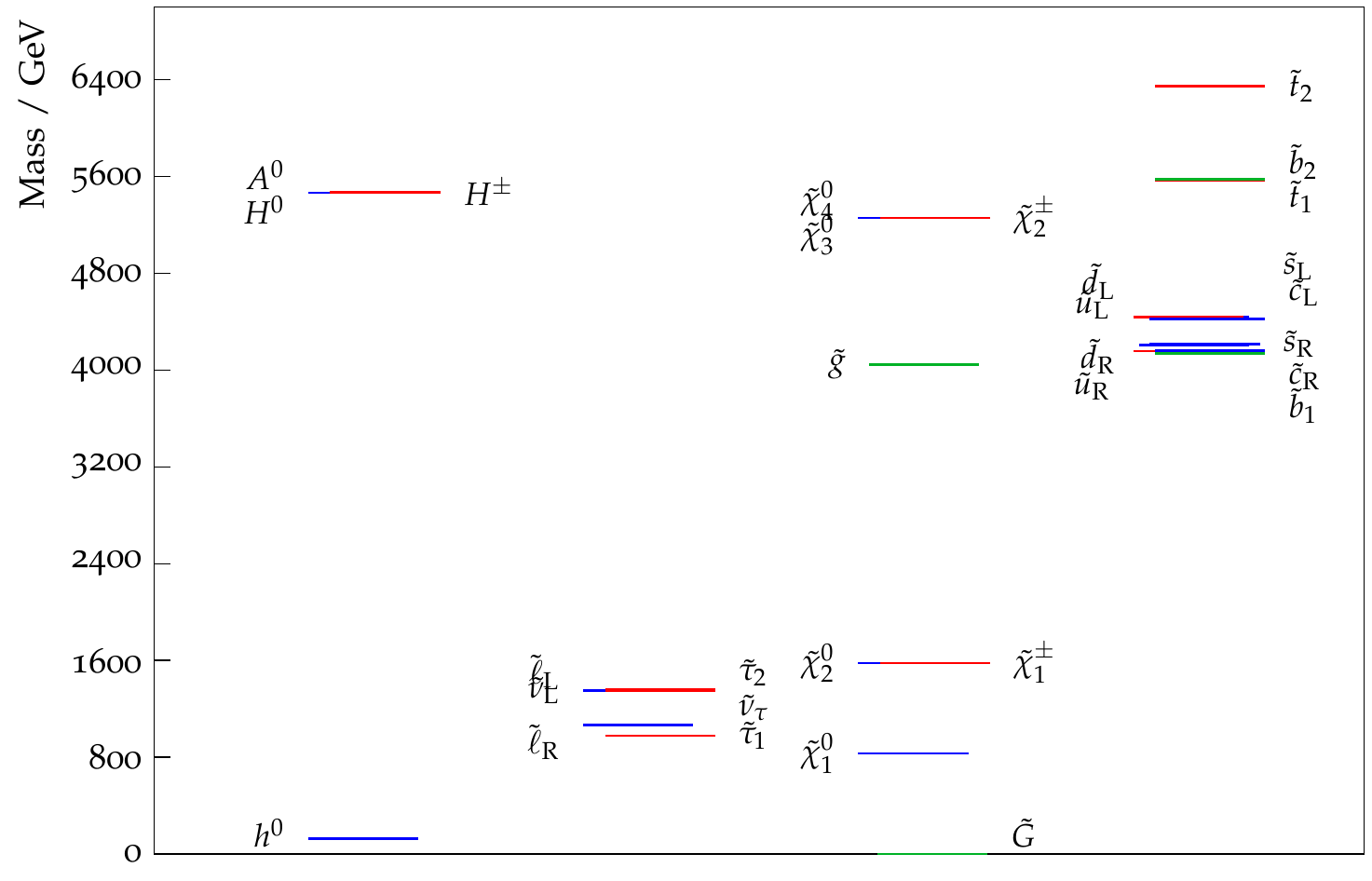}
   \end{subfigure}
   \centering
      \caption{Mass spectra for $M_{\text{Mess}}= 1\times10^{6}$ GeV (both sides), $\tan\beta=10$ and $\beta_\epsilon=0.01$ (left) and $\beta_\epsilon=0.05$ (right).  In each case, $\Lambda$ is fixed by the Higgs mass constraint.}
   \label{fig:LowMessSmallEps}
\end{figure}

\begin{figure}
\begin{subfigure}
   \centering
   \includegraphics[width = 0.4\textwidth]{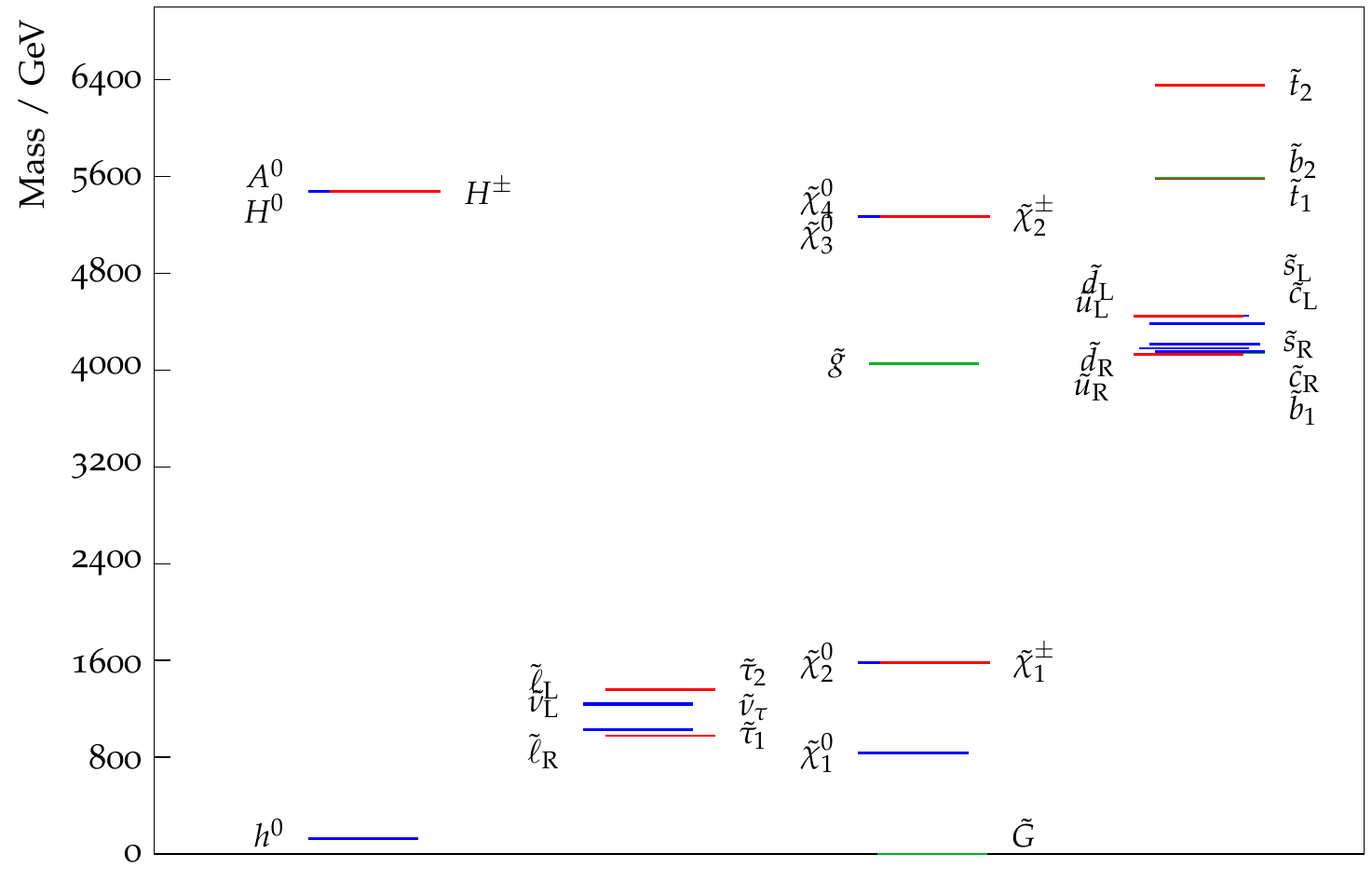}
   \end{subfigure}
 \begin{subfigure}
  \centering
   \includegraphics[width = 0.4\textwidth]{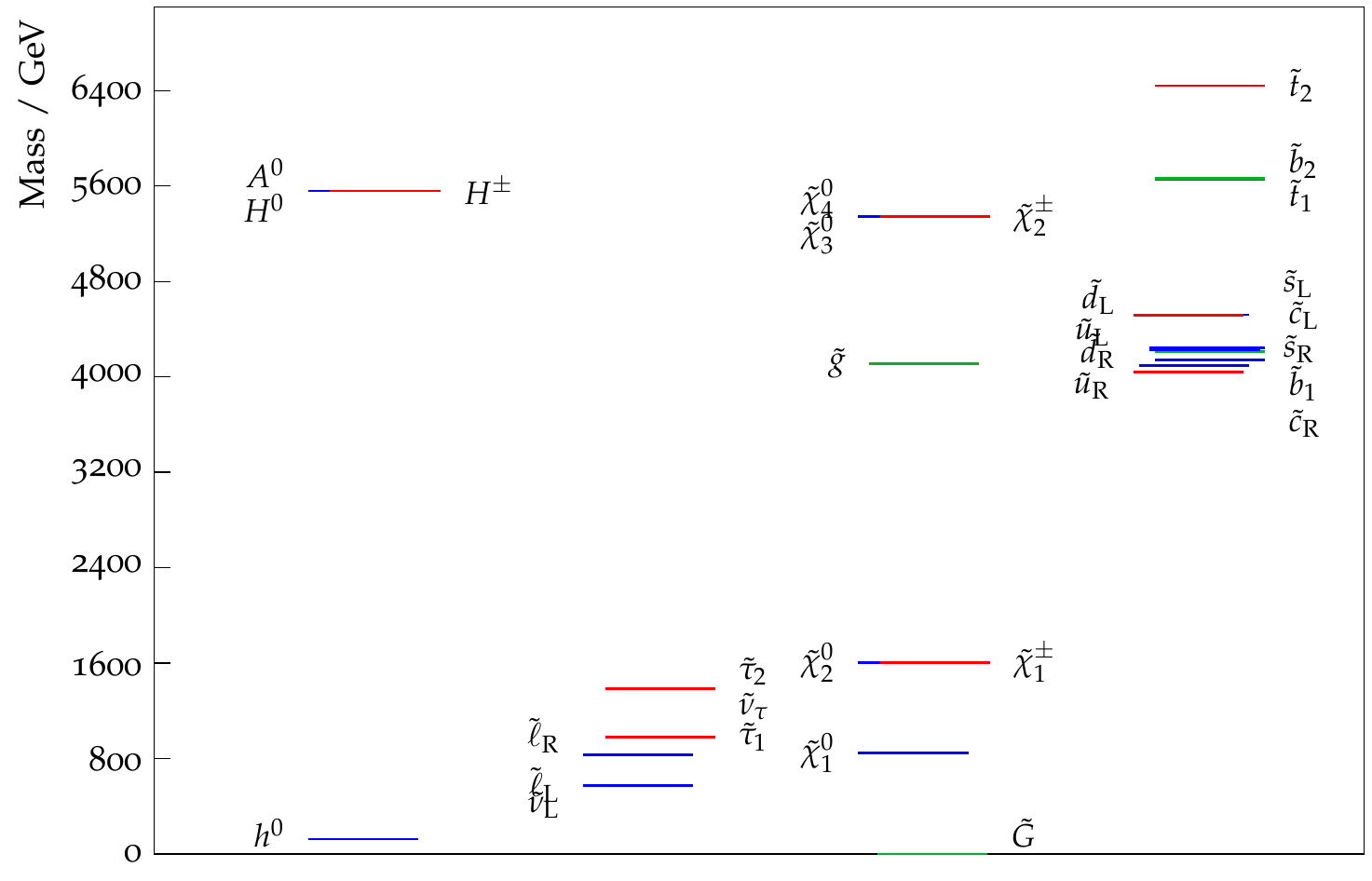}
   \end{subfigure}
   \centering
      \caption{Mass spectra for $M_{\text{Mess}}= 1\times10^{6}$ GeV (both sides), $\tan\beta=10$ and $\beta_\epsilon=0.1$ (left) and $\beta_\epsilon=0.2$ (right).  In each case, $\Lambda$ is fixed by the Higgs mass constraint.}
   \label{fig:LowMessBigEps}
\end{figure}
In Figures \ref{fig:LowMessSmallEps} and \ref{fig:LowMessBigEps}, we consider a smaller messenger mass of $M_{\rm Mess}=1\times10^6$ GeV. Displayed on the left-hand side of Figure \ref{fig:LowMessSmallEps} is the superpartner mass spectrum for this messenger mass, with $\beta_\epsilon=0.01.$ As seen previously in \cite{Everett:2018wrn}, a lower messenger scale leads to a large mass spectrum for fixed value of $\tan\beta$, due to the smaller size of the stop mixing. Since we choose $\Lambda$ such that $m_h\simeq 125$ GeV, a low messenger mass necessitates a larger value of $\Lambda$, and therefore leads to a heavier spectrum. The squark masses are no longer demarcated into two distinct groupings, but rather split between 4 and 6.4 TeV. The lightest squark is a sbottom, while the heaviest is a stop. There are four major squark groups. In decreasing mass order they are: the heaviest stop, the lighter stop and heavier sbottom, the left-handed squarks in generations one and two, and lastly the right-handed first and second generation squarks along with the other sbottom. The NSLP in this scenario is a bino-like neutralino.

The right-hand side of Figure \ref{fig:LowMessSmallEps} shows the spectrum for $\beta_\epsilon=0.05.$ Much like Figure \ref{fig:HighMessLowEps}, the increase in $\beta_\epsilon$ pushes the masses of the right-handed sdown and sup down, as well as the lightest left-handed charged slepton and sneutrino. The lightest squark continues to be a sbottom, but one can see the splitting amongst the masses of the lighter squarks begin to take shape. If now turn to the left-hand side of Figure \ref{fig:LowMessBigEps}, where $\beta_\epsilon=0.1,$ we see that the the general behavior as seen in the previous three spectra for $M_{\rm Mess}=1\times10^6$ GeV continues. What is new, however, is that the lightest squark is now a right-handed sup, much like was the case for the messenger scale $M_{\rm Mess}=1\times10^{12}$ GeV. We see that the lighter squark masses continue to split. 
Lastly, the right-hand side of Figure \ref{fig:LowMessBigEps} exhibits new behavior as compared to the spectra for a higher messenger mass. For example, the NSLP is left-handed slepton, as opposed to a neutralino.

%We pause here to comment on issues of perturbativity and consistency of the truncation of the nonrenormalizable operators to the single operator given in Eq.~(\ref{nonrnoperator}). As we have seen, several superpartner masses get pushed lower for larger values of $\beta_\epsilon.$ The question is up to what value of $\beta_\epsilon$ can we still consider the operator in Eq. (\ref{nonrnoperator}) a small correction to the leading order structure. If one increases $\beta_\epsilon$ enough, the left-handed sleptons become tachyonic, but by that point $\beta_\epsilon$ exceeds $0.5$.  Note that for $\beta_\epsilon \rightarrow 1$, the truncation of the set of nonrenormalizable operators to the single operator of Eq.~(\ref{nonrnoperator}).  This remains to be determined. (**LE comment: I believe this is unnecessary since we tie the beta_epsilon's to the second generation SM charged fermion masses.

We now find it instructive to investigate the behavior of this model over a wider range of $\Lambda$ and messenger mass. In Figures \ref{fig:LambdaSleptonSmaller} and \ref{fig:LambdaSleptonLarger}, we plot the predicted Higgs mass (solid contours), lightest slepton mass (dotted contours) and right-handed sup mass as $\Lambda$ and $M_{\rm Mess}$ are varied. We do this for four different values of $\beta_\epsilon.$ We see that for a phenemonologically viable point of parameter space (i.e $m_H=125$ GeV), the mass of the lightest slepton decreases. Eventually, there are points in $(\Lambda,M_{\rm Mess})$ parameter space that both provide a viable Higgs mass, and predict a slepton NLSP of less than 1 TeV.

\begin{figure}
\begin{subfigure}
   \centering
   \includegraphics[width = 0.4\textwidth]{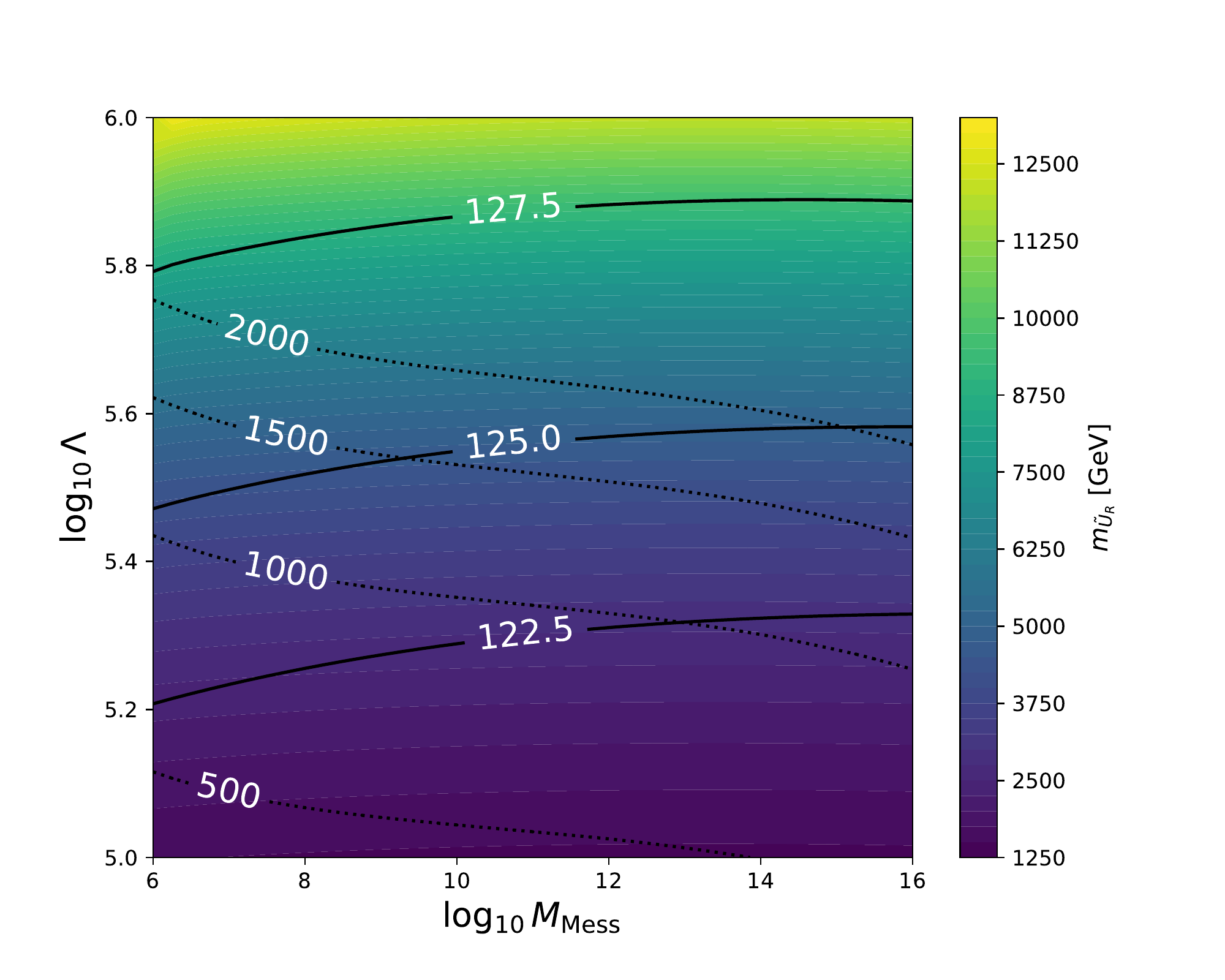}
   \end{subfigure}
 \begin{subfigure}
  \centering
   \includegraphics[width = 0.4\textwidth]{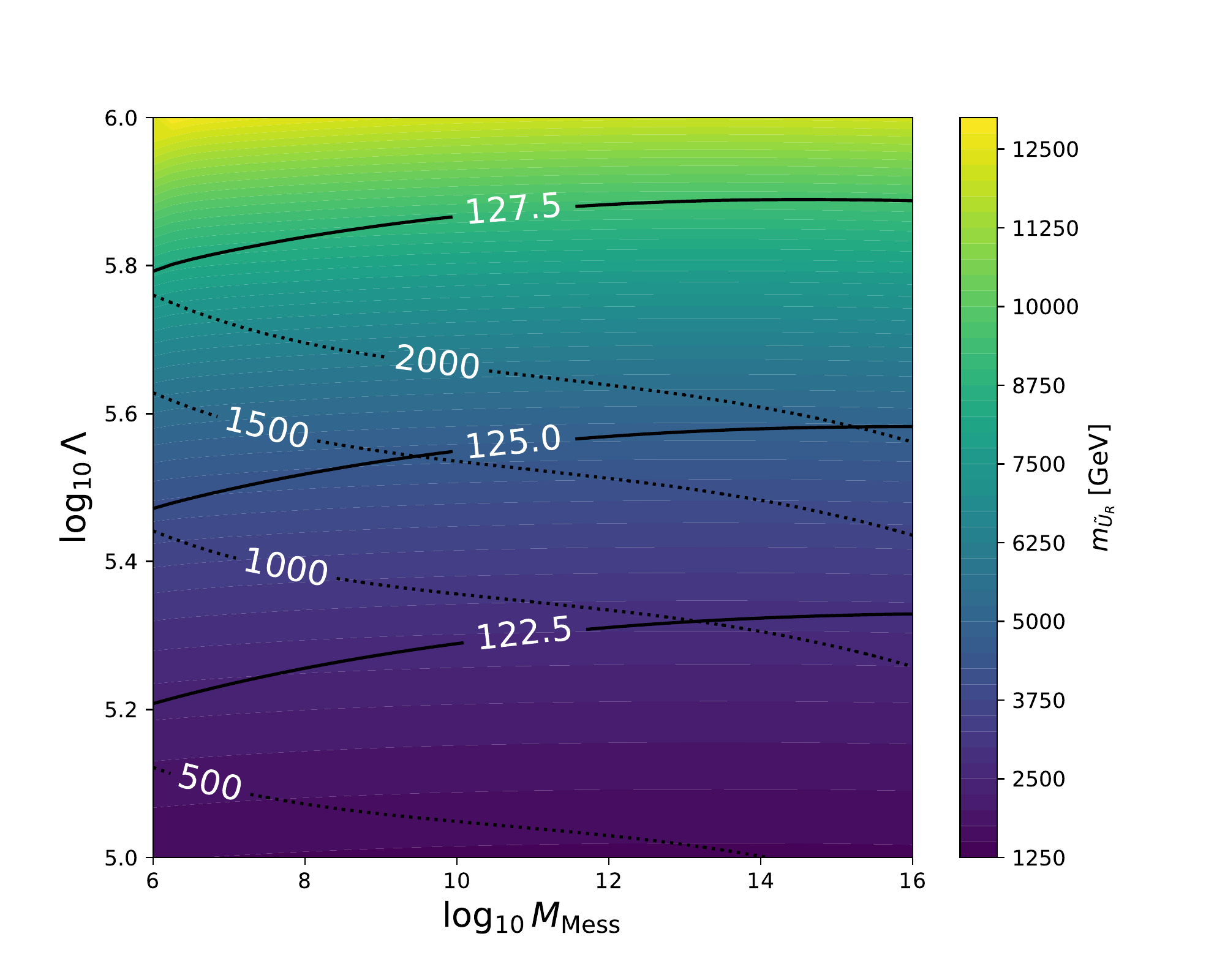}
   \end{subfigure}
   \centering
      \caption{The Higgs mass (solid contours), right-handed sup mass (color shading) and right-handed selectron masses (dotted contours) in this scenario with $\beta_\epsilon=0$ (left) and $\beta_\epsilon=0.05$ (right).}
   \label{fig:LambdaSleptonSmaller}
\end{figure}

\begin{figure}
\begin{subfigure}
   \centering
   \includegraphics[width = 0.4\textwidth]{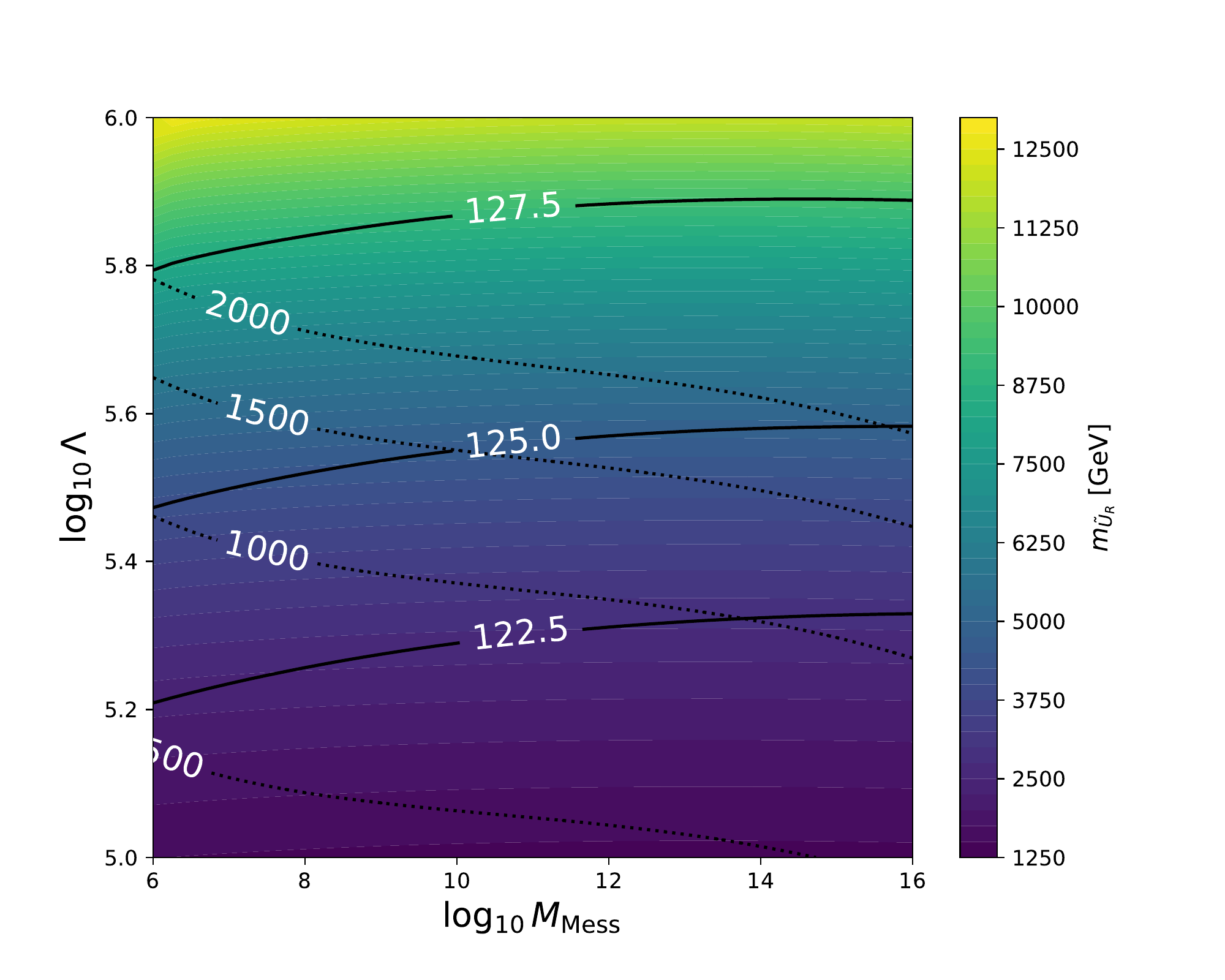}
   \end{subfigure}
 \begin{subfigure}
  \centering
   \includegraphics[width = 0.4\textwidth]{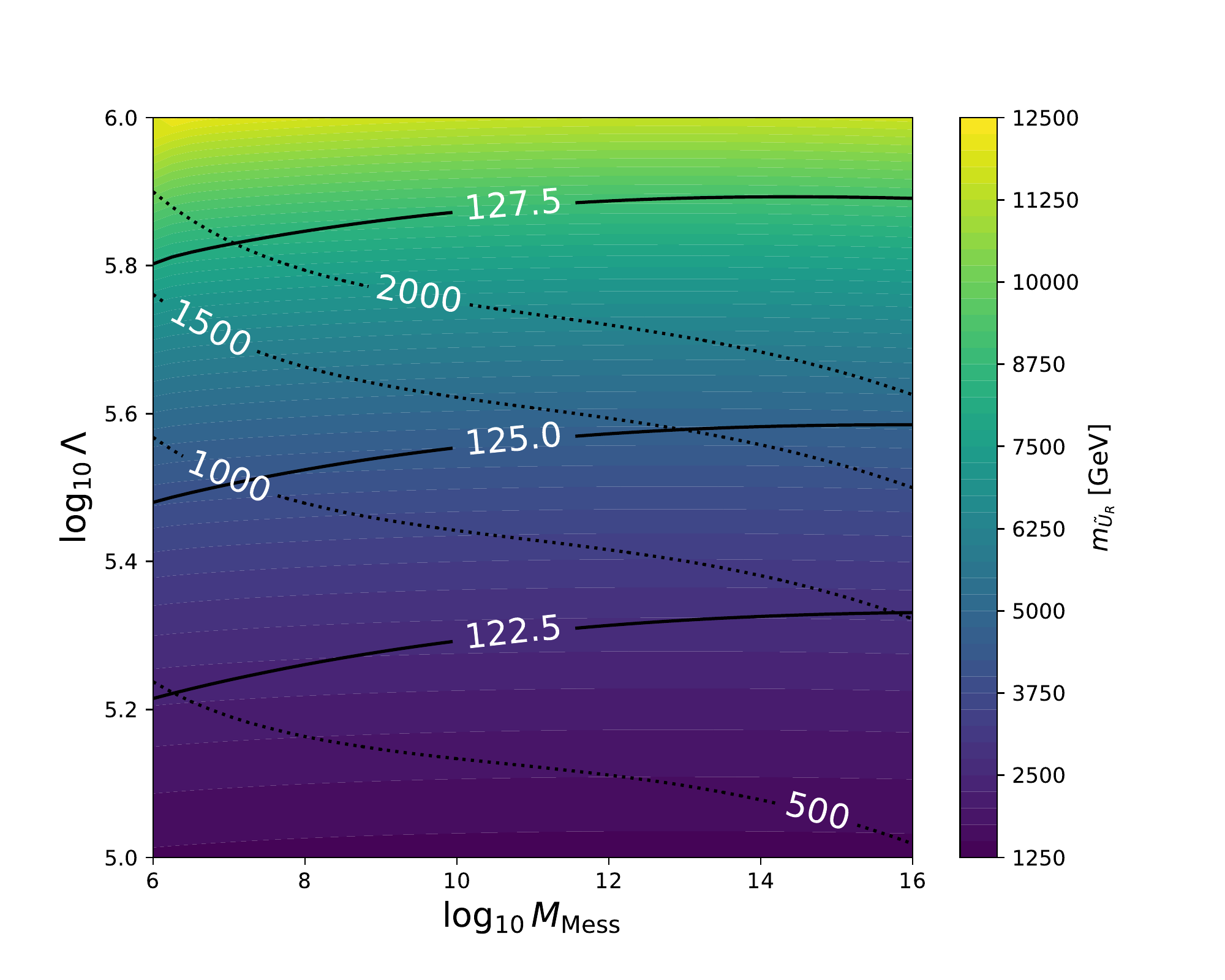}
   \end{subfigure}
   \centering
      \caption{The Higgs mass (solid contours), right-handed sup mass (color shading) and right-handed selectron masses (dotted contours) in this scenario with $\beta_\epsilon=0.1$ (left) and $\beta_\epsilon=0.2$ (right).}
   \label{fig:LambdaSleptonLarger}
\end{figure}

%\clearpage
\section{Conclusions}
In this paper, we have investigated the generation of fermion masses and quark mixing within a specific model of flavored gauge mediation in which the Higgs and messenger doublets are connected by the discrete non-Abelian symmetry described by $\mathcal{S}_3.$ This group also provides a framework for a partial family symmetry. This scenario requires the introduction of two messenger doublets that mix with the electroweak Higgs doublets via the $\mathcal{S}_3$ symmetry, rendering it an effective $N=2$ gauge mediation model with messenger Yukawa corrections. The phenemenology of this scenario in the case that only MSSM Yukawa couplings at the renormalizable level were included, and only the third generation SM fermions had nonzero masses, was investigated in \cite{Everett:2016meb} and \cite{Everett:2018wrn}.

We build on those previous analyses with the introduction of a nonrenormalizable perturbation of the superpotential couplings, which generates a hierarchically smaller mass for the second generation SM fermions, and leaves the first generation massless. In this paper, we showed that with a judicious choice of the nonrenormalizable operator, mixing among the first and second generation can result, and a Cabibbo angle of an appropriate size was able to be generated. While the scenario generically results in the possibility of flavor-violating couplings, we show that in a specific limiting case of the model parameters, the resulting messenger Yukawas in the diagonal quark mass basis yield flavor-diagonal corrections to the soft supersymmetry parameters, resulting in a scenario with few input parameters. %The scenario in this specific limit has four continuous parameters that govern the superpartner masses: the messenger scale $M_{\rm Mess},$ the scale $\Lambda \sim F/M_{\rm Mess},$ $\tan\beta$ and $\beta_\epsilon,$ which characterizes the strength of the nonrenormalizable operator. The model also has one discrete parameter (the sign of $\mu$).
We see in this context that the superparticle spectra are at most $4-6$ TeV, with the distribution of sparticle masses within this range being affected by the strength of the non-renormalizable perturbation. This highly predictive model thus provides a window into TeV-scale supersymmetry. Furthermore, as this model generically introduces nontrivial flavor structure, we now have a starting point for more stringent tests of supersymmetric theories using precision flavor experiments.

\begin{acknowledgments}

L.L.E. is grateful to D.~J~H.~Chung and M.~McNanna for their helpful input at the early stages of  this work.  This work is supported by the U. S. Department of Energy under the contract DE-SC0017647.
%contracts DE-FG-02-95ER40896 and DE-SC0017647.
\end{acknowledgments}

\bibliographystyle{prsty}

\end{document}